\documentclass[12pt,twocolumns]{iopart}
\usepackage{isolatin1}
\usepackage{graphicx}

\begin{document}
\def\etal{\emph{et al }}
\def\ibid{\emph{ibid.}}

\review{Stress-corrosion mechanisms in silicate glasses}

\author{Matteo Ciccotti}

\address{Laboratoire des Colloïdes, Verres et Nanomatériaux, UMR
5587, CNRS, Université Montpellier 2, Montpellier, France}

\ead{matteo.ciccotti@univ-montp2.fr}

%\doublespacing

\begin{abstract}
The present review is intended to revisit the advances and debates in the comprehension of the mechanisms of subcritical crack propagation in silicate glasses almost a century after its initial developments. Glass has inspired the initial insights of Griffith into the origin of brittleness and the ensuing development of modern fracture mechanics. Yet, through the decades the real nature of the fundamental mechanisms of crack propagation in glass has escaped a clear comprehension which could gather general agreement on subtle problems such as the role of plasticity, the role of the glass composition, the environmental condition at the crack tip and its relation to the complex mechanisms of corrosion and leaching. The different processes are analysed here with a special focus on their relevant space and time scales in order to question their domain of action and their contribution in both the kinetic laws and the energetic aspects.
%200 words. One paragraph.
\end{abstract}

%\pacs{1315, 9440T}

\submitto{\JPD}

\maketitle

\tableofcontents

\newpage

\section{Introduction}

Silicate glasses are highly homogeneous and isotropic materials at scales larger than a few nanometres, thus making continuum descriptions well suited down to submicrometric scales. This strong degree of homogeneity is at the origin of the best glass properties, such as the excellent transparency, but it is also partly the cause of a major limitation: brittleness! Glass products break very easily when subjected to even weak impacts with hard materials (Mould 1967), and also the static stress that can be sustained by a glass product for a convenient time without fracturing is quite low (Preston 1942, Gy 2003).
The extreme homogeneity of glass along with the elevated strength of its covalent cohesive bonds cause a huge stress concentration in a nanometric region around small mechanical defects. The local evolution of damage is thus very efficient, involving very little bulk dissipation and relatively efficient conversion of the mechanical energy into the surface energy required to create new fracture surfaces.

The earliest work of Griffith (1921) has identified the origin of the low strength of glass to the unavoidable presence of small flaws in glass objects. These flaws are generally interpreted as submicrometric surface cracks that escape notice under optical inspection.
Worse than this, these stress concentrators are responsible for a local acceleration of the corrosive influence of the environment on the glass. The damage will then generally spread in time even under moderate global stresses, leading finally to a delayed failure, especially in the presence of a moist atmosphere. This phenomenon, named static fatigue, can be interpreted as the combination of an initial stage of differential stress-enhanced corrosion (Charles et Hillig 1962), leading to the progressive sharpening of the flaws until the crack tip radius reaches a molecular (nanometric) dimension, and a second stage of stationary slow crack growth (Wiederhorn and Bolz 1970) with the stress-corrosion reactions concentrated at the molecularly sharp crack tip.

The development of fracture mechanics has allowed a rationalisation of the analysis of the fracture resistance of glass by describing the conditions of propagation of a single major fracture in a glass sample. The earliest introduction by Griffith of the strain energy release rate $G$,
%describing the balance of the energetic conversion of mechanical energy into fracture surface energy,
and then of the stress intensity factor $K$ (Irwin and Washington 1957) have made it possible to determine a more fundamental fracture criterion and more robustly defined fracture parameters for crack propagation in the form of critical values $G_c$ or $K_c$. Yet, in most practical cases $G_c$ can be larger than twice the equilibrium surface energy of the material, implying the occurrence of other subsidiary dissipation mechanisms in the fracture process.

The brittleness of glass can thus be read as the elevated efficiency of conversion of elastic energy into surface energy, suggesting a relatively weak contribution of plastic deformation and other bulk dissipation mechanisms, which would involve at most a region of few nanometres around the crack tip. The values of $K_c$ for most glasses are of the order of 1 MPam$^{1/2}$, which is rather low, especially when compared to the 50 fold larger values in metals where the energy dissipation by bulk ductility is dominant in the fracture energy.

In the fracture mechanics framework the subcritical propagation by stress-corrosion can be well rationalised and represented as a $v(K)$ or $G(v)$ relation for $K<K_c$ (Wiederhorn and Bolz 1970). This was shown to generally involve three propagation regions (cf.\ Fig.\ \ref{fig:ThreeRegions}). Region I corresponds to the cited stress-corrosion regime, where the crack propagates due to a stress-enhanced corrosion reaction, which is strongly dependent on the environment. In region II the propagation velocity is limited by the transport kinetics of environmental corrosive molecules to the crack tip. In region III, where the stresses are strong enough to induce the bond breaking in the absence of corrosion reactions, the velocity rises very sharply with $K$, eventually reaching the critical propagation at $K_c$. For some glass compositions, a threshold behaviour appears for $K=K_0$ leading to what is generally called region 0 (Michalske 1977).

Phenomenological equations such as Wiederhorn's (cf.\ section \ref{sec:RegionI}) can describe the dependence of crack velocity on stress and on environmental parameters for most typical glasses and are compatible with a thorough consistent modelling based on the sharp-crack atomic-bonding paradigm (Lawn 1993). Yet the detailed nature of the stress-corrosion mechanisms that occur at the crack tip have been debated for decades (Marsh 1964b; Maugis 1985; Gehrke \etal 1991; Tomozawa 1996), and a general disagreement can be found on the relevance of several accessory phenomena that may participate in the stress-corrosion mechanisms at different stages of the process. Stress-corrosion can involve a complex interplay between the diffusion of reactive molecules (mainly water) into the crack cavity and into the glass network, the corrosion (or dissolution) of the network itself, and the migration of weakly bonded alkali ions under chemical or stress gradient (Gehrke \etal 1991; Bunker 1994). All these phenomena are typically very slow under ambient conditions in the unstressed material, but they can significantly accelerate in the highly stressed neighbourhood of the crack tip, depending especially on the nature of the environment and of its confinement. Their time scales will be progressively accelerated in a series of smaller and smaller shells surrounding the crack tip (cf.\ Fig.\ \ref{fig:MechSchemeA}), defined by the stress being larger than some critical value.
However, since the crack tip is propagating, the different phenomena will be able or not to spread in each shell depending on a competition between their time scale in each shell and the crack propagation velocity.  Since the size of the activated shells can vary from microns to nanometres depending on the interplay between space and time scales, the modelling of these phenomena (which will be reviewed in more detail in next sections) requires nanometre scale resolved investigation techniques and questioning about the relevant physical laws at the nanoscale. Such studies should be very promising to solve the debates by direct observation and to relate the phenomenological parameters to the specific composition and structure of glasses.

The development of several nanoscale investigation techniques in the 80's has led to significant insights into the comprehension of the different mechanisms (cf.\ sections \ref{sec:plasticity} to \ref{sec:surfaces}). In particular SEM, TEM and AFM measurements of both {\it post-mortem} crack surfaces and direct {\it in-situ} observations of the crack tip neighbourhood have permitted the investigation of the space and time scales of occurrence of these phenomena under specific conditions. The progressive increase in resolution down to the micron scale of several structural and compositional analysis techniques (such as Raman, IR, Brillouin spectroscopies, X-Ray, electron and neutron scattering, NMR, XPS, SIMS) have also permitted investigations of the alterations of the bulk glass near the crack tip or at freshly fractured surfaces.
These techniques, combined with the increasing power of molecular and quantum dynamics simulations permit great insights into the understanding of the combination between mechanical and chemical processes acting at the crack tip.

Section \ref{sec:SpaceTime} will first focus on some relevant concepts of fracture mechanics. In section \ref{sec:Kinetics} the most solidly established features of the slow crack propagation kinetics in glasses will be described
and interpreted in the framework of the classic sharp-crack atomic-bonding paradigm proposed in the '70s.
The section \ref{sec:Insight} of the review will present a deeper discussion of the relations between the crack propagation and the chemical mechanisms of stress-corrosion, along with a critical analysis of the points which are still controversial and of the recent experimental evidences and efforts which have been done to make them clearer. A concluding section will discuss the perspectives and promises of the development of this interesting research field.

\section{Space, time and energy scales of fracture mechanics}
\label{sec:SpaceTime}

The world of fracture mechanics is extremely wide and rich of theories and variants that account for the many peculiar manifestations in different materials. It is difficult to figure out the `one theory' that can encompass different phenomena such as atomic bond breaking, dislocation motion, viscoelastic-plastic flow, damage spreading,  corrosion, ion exchange, cavitation, fingering and a whole plethora of other fancy effects. Different application communities generally end up with following different paths, along with different terms, variables and philosophies. Luckily enough, the classic Linear Elastic Fracture Mechanics (LEFM) was mainly developed to tackle strength problems of brittle materials, and glass has often been a privileged material for its understanding and application (cf.\ Lawn, 1993). In the present section, I will revisit the main relevant concepts and theories of fracture mechanics and propose a synthetic point of view that will guide the analysis in the present review by focusing on the space, time and energy scales of the different physico-chemical phenomena occurring during fracture.

\begin{figure}[!h]
\centering
\begin{minipage}[t]{0.48\linewidth}
   \centering
   \includegraphics[width=7 cm]{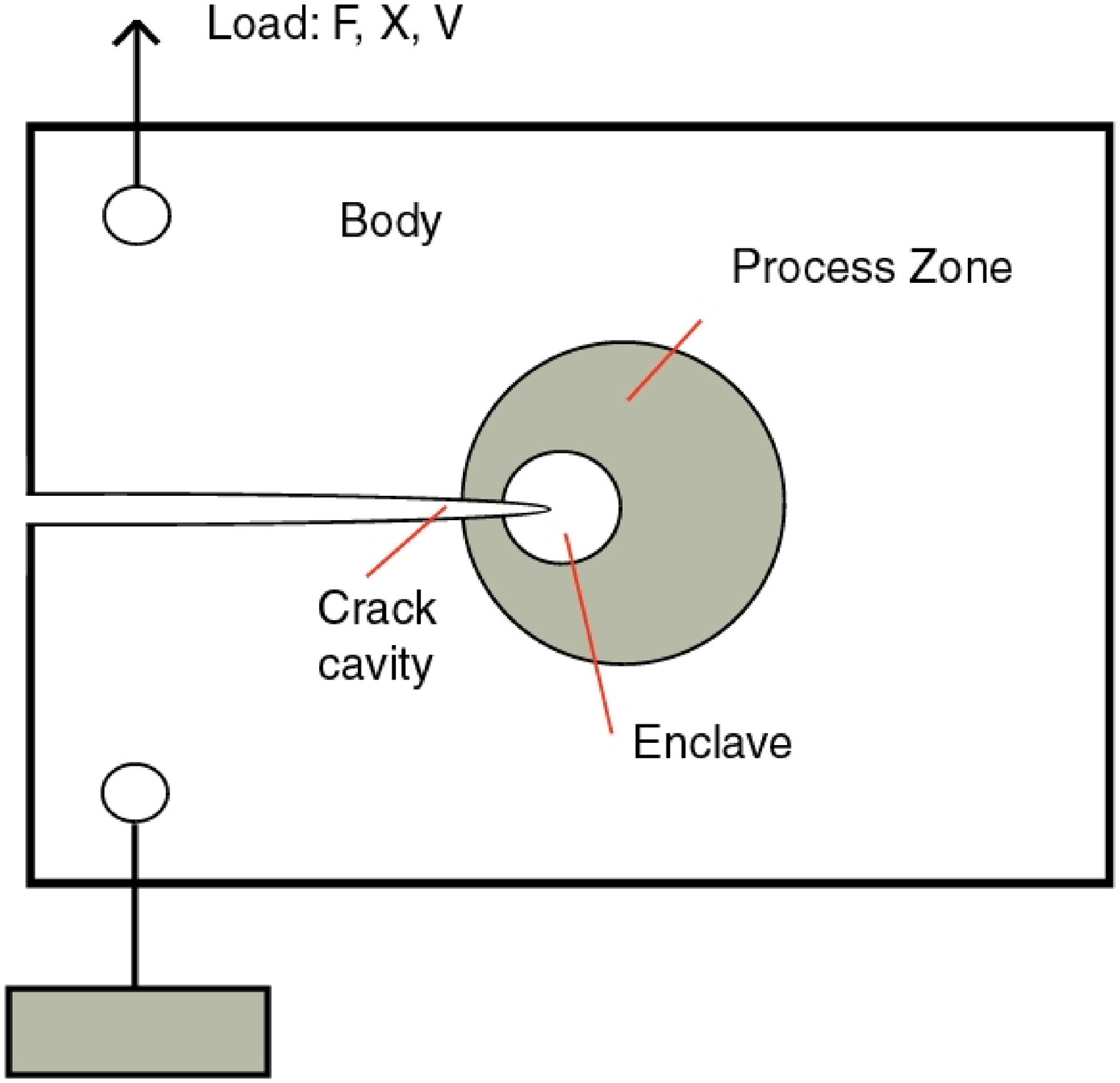}
   \caption{Graphic representation of the mechanical elements for fracture mechanics.}
   \label{fig:MechSchemeA}
\end{minipage}%
\hfill
\begin{minipage}[t]{0.48\linewidth}
   \centering
   \includegraphics[width=7 cm]{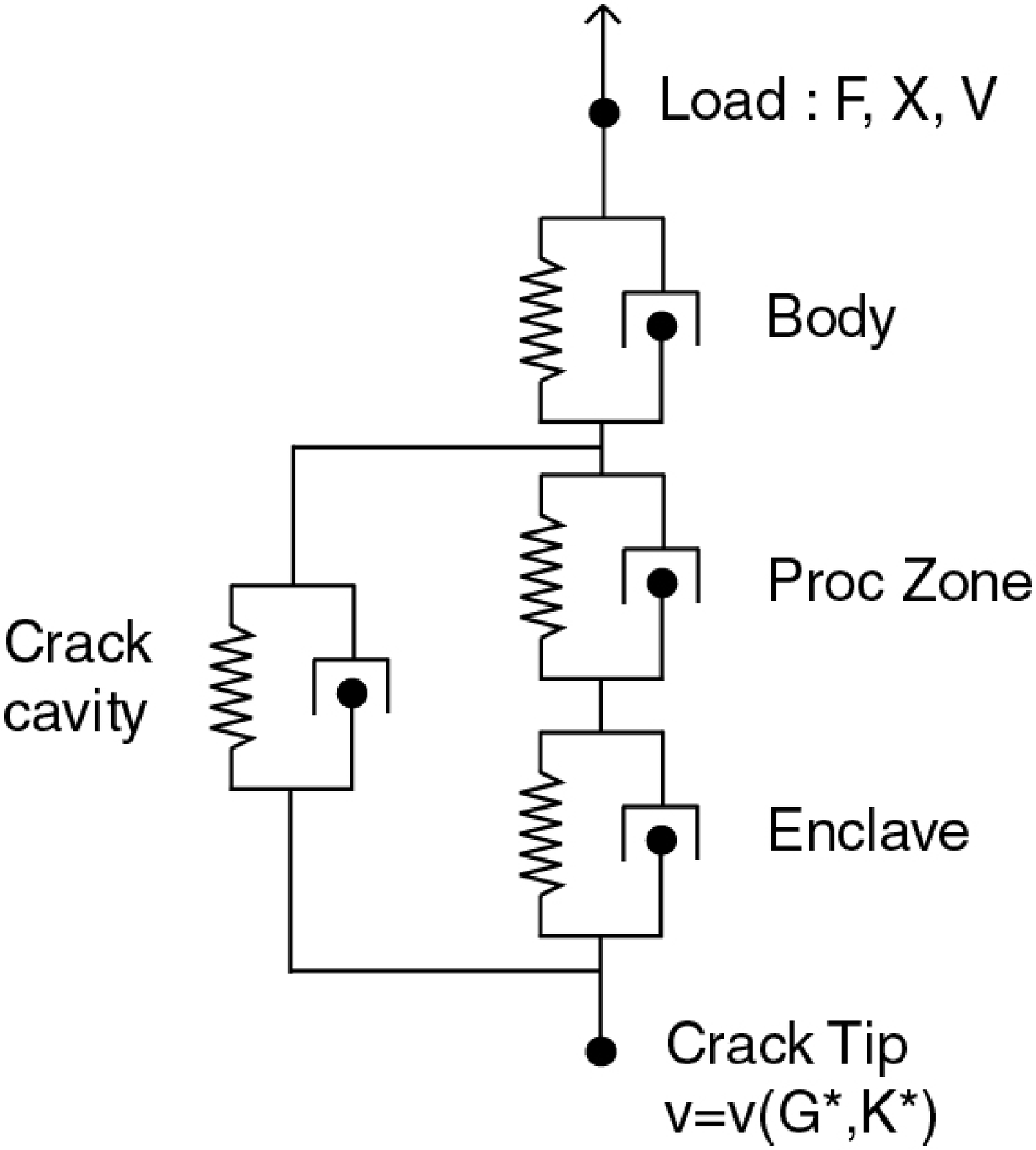}
   \caption{Symbolic block representation of the different mechanical elements.}
   \label{fig:MechSchemeB}
\end{minipage}
\end{figure}

Although the ideal basic mechanism for fracture is the conversion of mechanical energy into surface energy (Griffith 1921), fracture propagation is generally complicated by the activation of subsidiary dissipation mechanisms that generally act in a bulk neighbourhood of the crack tip called the process zone (Irwin 1960), but that can also extend to the whole material volume or act between the open crack walls as illustrated in the schematic model in Fig. \ref{fig:MechSchemeA}. The energy dissipated by each of these mechanisms has to be taken into account to estimate the energy per unit area that is necessary to propagate the fracture.
To progressively account for nonlinear effects, irreversibility and dissipative processes, fracture mechanics has undergone several profound developments, which however were managed with appreciable devotion in order to preserve the formalism of LEFM (Orowan 1955; Irwin 1960; Barenblatt 1962; Rice 1968; Rice 1978; Lawn 1993). This is based on one hand on the use of surface energy balance related variables such as the strain energy release rate $G$ and on the other hand on variables describing the fundamental inverse square root dependence of the stress field on the crack tip distance such as the stress intensity factor $K$ (Irwin and Washington 1957)\footnote{The present review is limited to mode I tensile fracture for simplicity, the variable $K$ will stand for $K_I$}. Moreover, the two kinds of variable remain in general related by an equivalence in the typical form $G = K^2/E'$ ($E'$ being an equivalent Young's modulus) and the criteria for equilibrium and crack propagation can still generally be expressed by equations of the kind $G \geq G_c$ or $K \geq K_c$ by only changing the exact definition of the variables used (Lawn 1993).

As concerns the description of the kinetics of fracture, we can distinguish two main families of approaches: the ones where the dissipated energy dominates the surface energy and the ones where the crack tip debonding processes play the dominant role. These approaches involve the use of different variables and diagrams, but also reflect different philosophies.
In order to provide a common frame it is useful to analyse the mechanical scheme in Fig.\ \ref{fig:MechSchemeA}. In a typical fracture propagation experiment, the external load can be represented by a given force $F$, displacement $X$ or traction velocity $V$ applied to a loading point. In the most general case, the test sample can be subdivided into a series of regions that progressively transfer the energy towards the crack tip, such as the body, the process zone (one or a series of shells depending on the number of potential dissipative mechanisms), a local near-tip elastic enclave, and the region inside the crack cavity. The mechanical role of each of these regions is schematically represented in the block diagram in Fig.\ \ref{fig:MechSchemeB}, where the nature and spatial extension of each block can change between different materials and depend on the time scales of loading and crack propagation, each block possibly being elastic, viscoelastic, viscoplastic, etc (the symbols used in the graph are just evocative!). In order for $K$ to be definable and the LEFM to apply, the body block should be purely elastic and the process zone size should be small with respect to the sample and fracture length sizes. According to the theory for brittle materials (Lawn 1983), it is possible to define a small enclave zone near the crack tip that behaves in an essentially elastic way and that allows the definition of the local values of $G^*$ and $K^*$ that are relevant for describing crack tip processes.

\begin{figure}[!h]
\centering
\begin{minipage}[t]{0.48\linewidth}
   \centering
   \includegraphics[width=7 cm]{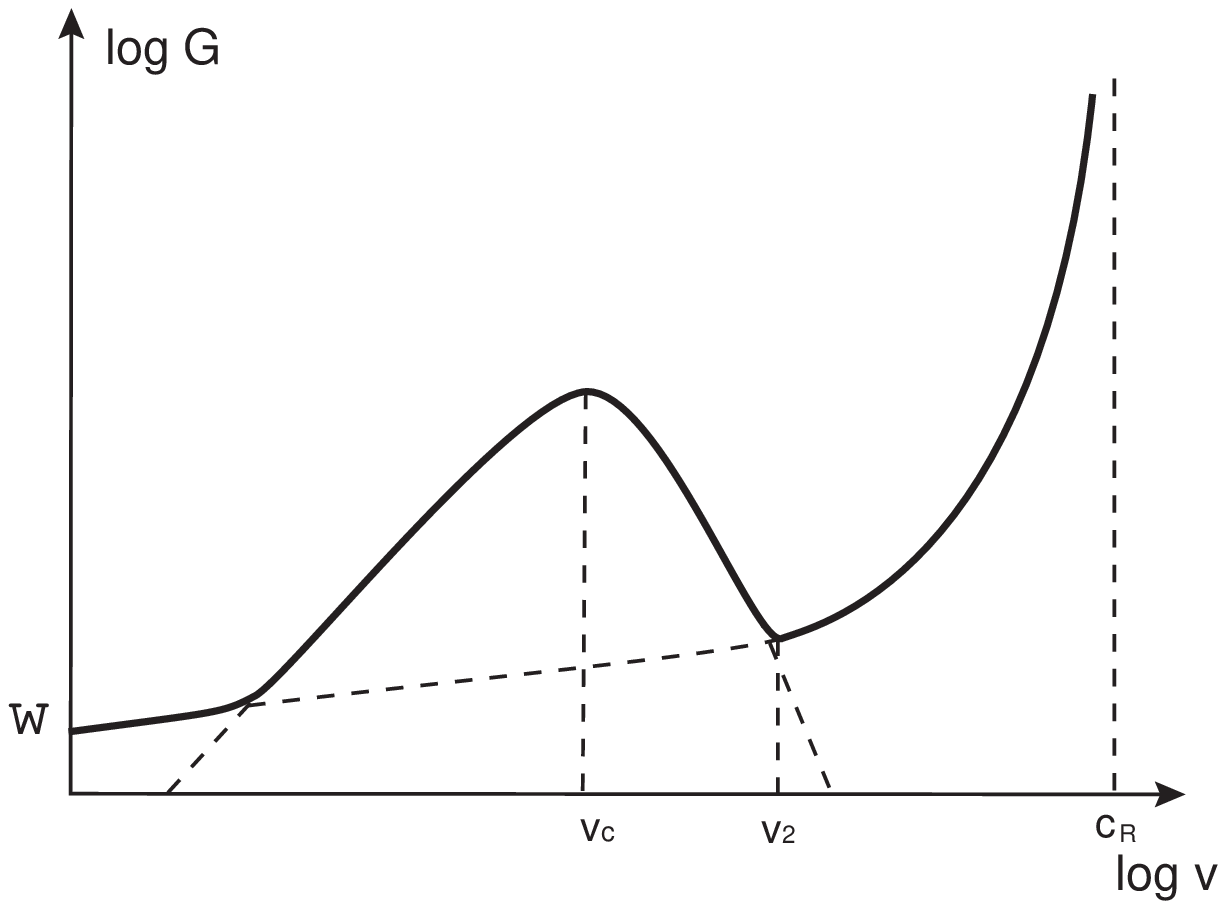}
   \caption{Typical $G(v)$ diagram for crack propagation in polymers and adhesives ($w = G_0$).}
   \label{fig:MaugisGv}
\end{minipage}%
\hfill
\begin{minipage}[t]{0.48\linewidth}
   \centering
   \includegraphics[width=5 cm]{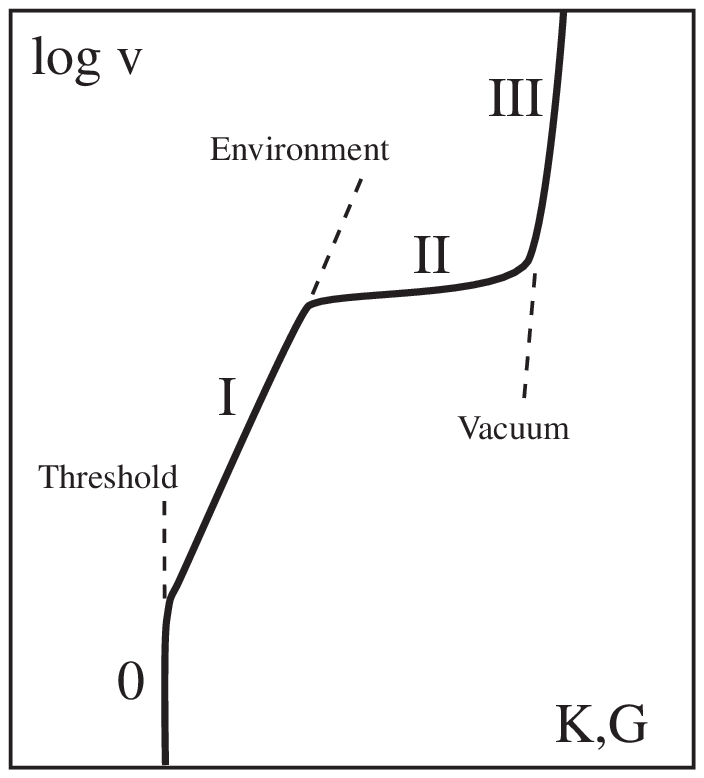}
   \caption{Typical $v(K)$ or $v(G)$ diagram for subcritical crack propagation in glass.}
   \label{fig:ThreeRegions}
\end{minipage}
\end{figure}

In the first approach (mainly formalised by Maugis (1985) to describe fracture in polymers and peeling of adhesives) fracture propagation in a given material is generally characterised by $G(v)$ curves, representing the strain energy release rate as a function of the crack velocity (Fig.\ \ref{fig:MaugisGv}). This makes it particularly easy to consider the energy balance of fracture propagation in relation to the time scales of deformation solicited by different crack propagation velocities. In this formalism, the low velocity limit can be identified with the `equilibrium' surface energy $G_0=2\gamma_0$. Then the fracture energy increases with velocity due to enhanced activation of concomitant irreversible dissipative processes, which determine a stationary propagation curve in the form $G(v)$. For the case of viscoelastic materials, the slow fracture region is generally in the form of a power law of the kind $G(v) = G_0(1+A_Tv^n)$, where $A_T$ includes the temperature dependence. The increasing portion of the $G(v)$ curve eventually ends at some critical velocity $v_c$, where the dissipative process starts to be less and less effective,
since the kinetic scale becomes faster than the viscoelastic relaxation. For higher velocity the fracture energy increases again due to the manifestation of dynamic effects (Freund 1990) eventually leading to a divergence when the crack velocity approaches the sound propagation velocity\footnote{The observed limit propagation velocity is generally only a fraction of the sound velocity due to the occurrence of other kinds of dynamical instabilities (cf. the review of Fineberg and Marder 1999).}. The presence of a maximum in the $G(v)$ curve leads to an intrinsic mechanical instability in the fracture propagation dynamics for a critical value $G_c$ that should not be confused with the threshold surface energy $G_0=2\gamma_0$, although it is generally proportional to it. The range of velocities corresponding to a negative slope in the $G(v)$ curve does not allow stationary crack propagation and generally leads to stick-slip dynamics (cf.\ Ciccotti \etal 2004). The instability at $G_c$, i.e.\ the jump from slow to rapid fracture for a constant applied force, is of particular interest since it is marked by a sharp reduction of the amount of energy dissipated by fracture propagation, letting all the excess furnished energy to be converted to kinetic energy with the typical emission of acoustic bursts (Barquins and Ciccotti 1997).

In the second approach (mainly formalised in the work of Wiederhorn \etal (1982) on glass and Lawn (1993) for brittle materials) the focus is on the kinetics of a crack tip debonding process as a function of the applied stresses. The fracture propagation curves are generally represented by $v(K)$ curves, where the crack propagation velocity is considered as a variable dependent on the stress intensity factor (Fig.\ \ref{fig:ThreeRegions}). The global energy balance is somehow in a subsidiary position, and the fracture kinetics is determined by the rate of thermal activation of a generic discrete bond breaking reaction (Thomson \etal 1971) whose activation energy is reduced by the application of stresses, leading to an Arrhenius like exponential equation $v = v_0 \exp[-(\Delta E_a-bK)/k_BT]$, that may also be expressed in terms of $G$ (cf.\ section \ref{sec:RegionI}).
In this interpretation, the critical value $K_c$ for the transition to dynamic propagation should correspond to the end of the lattice trapped regime, while the equilibrium surface energy $2\gamma_0$ should correspond to a propagation threshold. % (cf.\ section \ref{sec:threshold}).
However, this basic modelling can only be argued for a single debonding mechanism. The inert fracture (cf.\ section  \ref{sec:RegionIII}) should thus be modelled separately from chemical assisted propagation (cf.\ sections \ref{sec:RegionI} and \ref{sec:threshold}).
We should however retain the idea that the position of the critical value $K_c$ is related to the discreteness and heterogeneity of the material rather than to a surface energy, since it represents the moment where the energy barriers for individual bond breaking tend to zero. The surface energies should rather be compared with the propagation thresholds for each mechanism. In materials which are heterogeneous on a large scale, this approach can be reformulated in terms of a larger basic grain size (Broberg 2003).

The two approaches can be usefully combined in brittle materials presenting a non negligible process zone, but where a small near-tip elastic enclave can be defined. To do that we should first assume that the local values of $G^*$ and $K^*$ in the enclave are effective in controlling the crack tip propagation kinetics according to a relation $v(K^*)$ (Lawn 1983). In the absence of forces inside the crack cavity, the energy dissipated in the process zone can be evaluated by the difference $G-G^*$ (applying the $J$ integral formalism of Rice 1968) and will cause both $G^*$ and $K^*$ to be smaller than their macroscopic counterpart $G$ and $K$. The effect of energy dissipation will thus translate into a shielding (or local relaxation) of the crack tip stresses. The presence of generalised forces in a limited portion of the near-tip crack cavity will also act as a shielding factor at the crack tip, and it may or may not involve energy dissipation depending on the nature of the interaction.

When describing the macroscopic fracture properties of a material, it is the relation of the external variables $G,K$ which is studied as a function of $v$. Even if assuming a uniquely defined kinetic law $v(G^*,K^*)$, these local variables remain hidden, and the presence of a process zone will cause the non-uniqueness of the macroscopic fracture laws $v(G,K)$. This induces a time- or crack-length dependent behaviour (R-curve) that may lead into controversial interpretation (Lawn 1993). These types of modelling will be shown later to be particularly effective in rationalising the propagation threshold behaviour (cf.\ section \ref{sec:threshold}).

As a concluding remark to this section,
it is generally useful to identify the different space, time and energy scales of each symbolic block (cf.\ Fig.\ \ref{fig:MechSchemeB}) in order to estimate if the different dissipation mechanisms will have time to activate and how they will influence the energetic balance or the fracture propagation laws. Several spatial scales are present in the problem, such as the sample lateral dimensions and thickness, the loading displacement, the crack length, the basic grain size of the material (molecular rings for glass), the size of the process zone (or of each shell of a multiple process zone), the enclave size, the length of the one or several cohesive zones in the crack cavity. Typical time scales are given by the loading or displacement rate, the crack velocity, the stimulation time
of each equivalent block and its characteristic relaxation time, the rate of thermal activation of different mechanisms, the rate of transport of different relevant chemical species in the bulk or in the crack cavity. We will try to keep these space and times scales in mind while proceeding in the description of the crack propagation kinetics in glasses and especially when trying to get a deeper insight into the stress-corrosion mechanisms.

\section{Crack propagation kinetics}
\label{sec:Kinetics}

A huge variety of possible oxide glasses may be created by melting together variable amounts of oxides, within specific compositional ranges that are strongly influenced by the quench speed (cf.\ Zarzycki 1991). %(Zachariasen 1932)).
The details of the chemical reactions and the competition between different propagation mechanisms will be altered, but many typical features of crack propagation remain substantially similar. The most influential difference is between glasses essentially made by network-former oxides ($SiO_2$, $GeO_2$, $B_2O_3$, $P_2O_5$, etc.) and glasses containing significant amounts of modifier oxides ($Na_2O$, $K_2O$, $CaO$, etc.). For that reason in the following discussion silica glass will mainly be described as a representative of the first class and a typical soda-lime glass will be used as paradigm to represent the second class of glasses.

The subcritical fracture propagation properties of glasses are generally studied using samples that allow the stable slow propagation of a single fracture such as the Double Torsion (DT, Evans 1972), the Dual Cantilever Beam (DCB, Srawley and Gross 1967) or the Double Cleavage Drilled Compression (DCDC, Janssen 1974). The measurement of the crack propagation velocity as a function of the applied stress intensity factor generally presents three (or four) characteristic regions, Fig.\ \ref{fig:ThreeRegions}.

\subsection{Region I: stress-corrosion regime}
\label{sec:RegionI}

\begin{figure}[!h]
\centering
\begin{minipage}[t]{0.48\linewidth}
   \centering
   \includegraphics[width=7 cm]{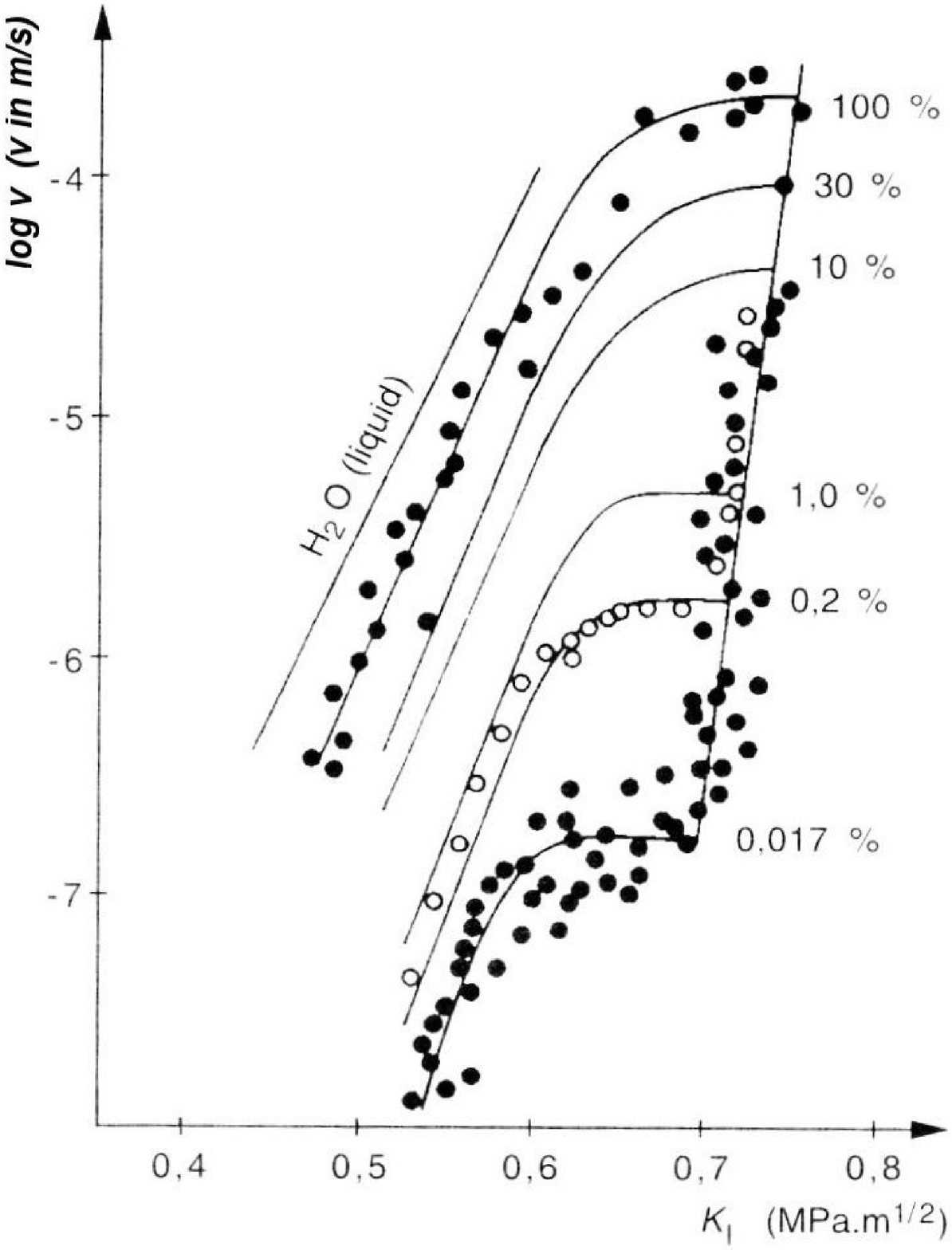}
   \caption{Effect of humidity on the crack propagation in soda-lime glass (from Wiederhorn 1967).}
   \label{fig:KVHGlass}
\end{minipage}%
\hfill
\begin{minipage}[t]{0.48\linewidth}
   \centering
   \includegraphics[width=7 cm]{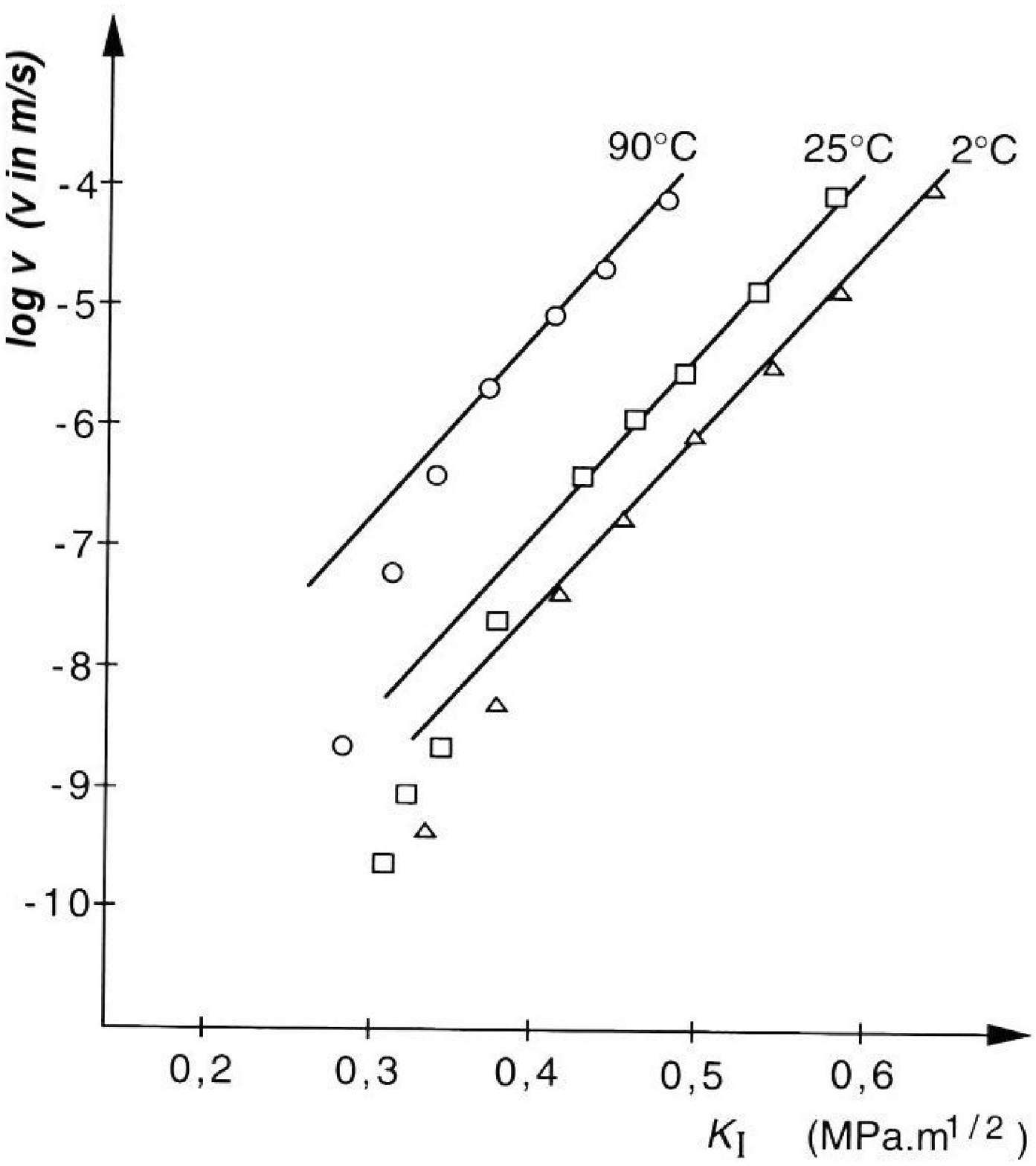}
   \caption{Effect of temperature on the crack propagation in soda-lime glass (from Wiederhorn and Bolz 1970).}
   \label{fig:KVTGlass}
\end{minipage}
\end{figure}

In region I, corresponding to the stress-corrosion regime, the crack propagation velocity is a strongly increasing function of the stress intensity factor $K$; it has an almost linear dependence on humidity in moist air (Fig.\ \ref{fig:KVHGlass}) and it increases with temperature (Fig.\ \ref{fig:KVTGlass}). All these dependencies can generally be fitted by the Wiederhorn (1967) equation:

\begin{equation}
\label{eq:Wiederhorn}
v^I = v_0 \exp(\alpha K)
  = A \left( {p_{H_2O} \over p_0} \right)^m \exp \left( - { \Delta E_a - b K \over RT} \right)
\end{equation}
where $p_{H_2O}$ is the partial pressure of the vapour phase in the atmosphere, $p_0$ is the total atmospheric pressure, and $R$ the gas constant; $A$, $m$, $\Delta E_a$ and $b$ are four adjustable parameters that take into account the dependence on the glass composition (cf.\ interpretation below).

However, due to the extremely strong dependence of velocity on the stress intensity factor $K$ and to the limited range of variation of $K$, the data in region I may also be fitted by a power law expression like in the Maugis (1985) approach:

\begin{equation}
v^I = v_0 (K/K_0)^n
\label{eq:PowerLaw}
\end{equation}
the exponent $n$ being generally between 12 and 50 for silicate glasses (Evans and Wiederhorn 1974). This form is by the way particularly useful (and indeed used) for developing simple analytical predictions of the life times in static and dynamic fatigue tests (Davidge \etal 1973). The difficulty in determining the most adequate relation results in a strong uncertainty in the predictions over very long periods.

For the same reasons, the region I data can also be fitted by a different exponential equation proposed by Pollet and Burns (1977) and applied to glasses by Lawn (1993):

\begin{equation}
\label{eq:PolletBruns}
v^I = v_0 \exp(\alpha G) = A \exp \left( - {\Delta E_a \over k_BT} \right) \exp \left( - {\alpha(G-G_0) \over k_BT} \right)
\end{equation}
where the energetic balance has a more central role, as will be discussed later concerning the threshold at $G_0$.

\begin{figure}[!h]
\centering
\begin{minipage}[t]{0.48\linewidth}
   \centering
   \includegraphics[height=9 cm]{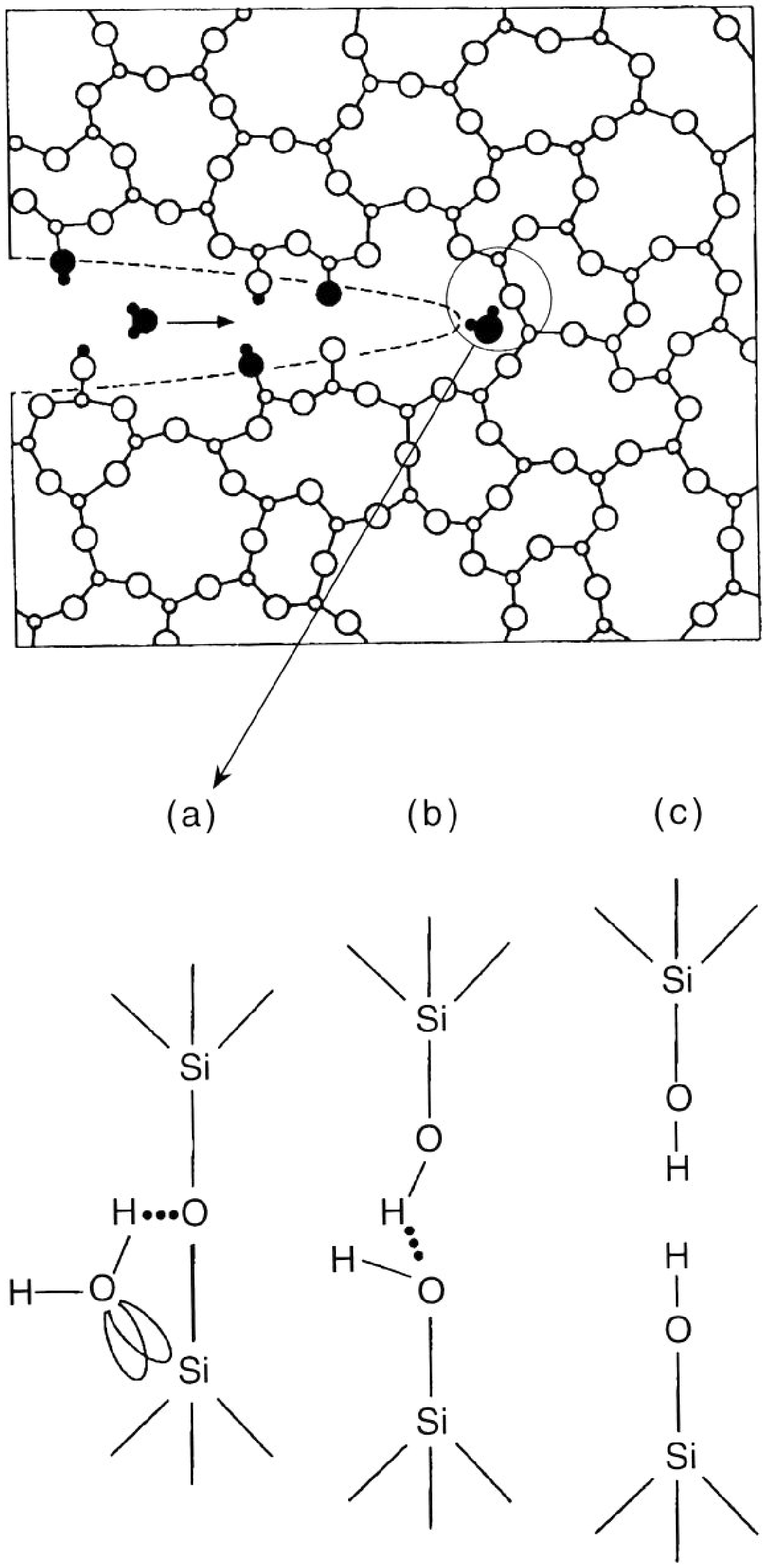}
   \caption{Basic mechanism of the stress-corrosion reaction (from Michalske and Freiman 1983).}
   \label{fig:SCSReaction}
\end{minipage}%
\hfill
\begin{minipage}[t]{0.48\linewidth}
   \centering
   \includegraphics[width=7 cm]{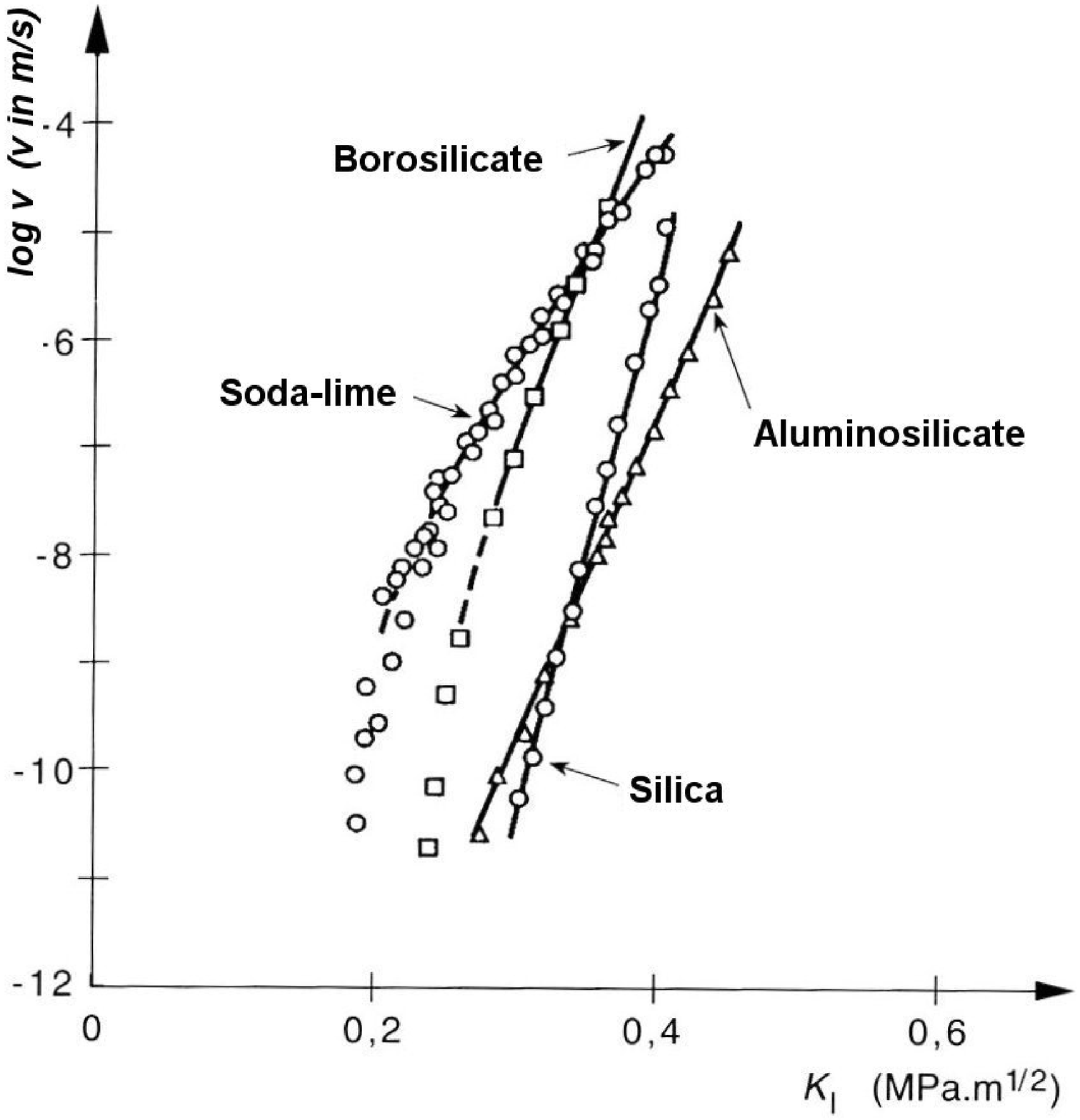}
   \caption{Effect of glass composition on the crack propagation in soda-lime glass (from Wiederhorn and Bolz 1970).}
   \label{fig:KVCGlass}
\end{minipage}
\end{figure}

The Wiederhorn equation (\ref{eq:Wiederhorn}) has been the most used in the description of region I in glasses since it provides a more direct interpretation of the chemical reactions between a glass of a given composition and the environment at the crack tip. In this framework, the basic mechanism of the stress-corrosion reaction was associated with the stress-enhanced thermal activation of a dissociative hydrolysis reaction represented in Fig.\ \ref{fig:SCSReaction} for the case of silica glass (Michalske and Freiman 1983).
Michalske and Bunker (1984) supported the positive effect of stress on the reaction rate by a molecular-orbital simulation of the
reaction on distorted siloxane bonds. They later provided experimental support for their theory by measuring the rate of hydrolysis of a series of strained silicate ring structures (Michalske and Bunker 1993).
For other glass compositions (Fig.\ \ref{fig:KVCGlass}), several other reactions are possible, but they share the similar feature of being stress-enhanced and thermally activated, thus explaining the exponential (Arrhenius like) term in equation (\ref{eq:Wiederhorn}), where $\Delta E_a$ can be interpreted as an activation enthalpy in the absence of stress and $b$ can be expressed in terms of an activation volume (of molecular dimensions) by relating the stress intensity factor to the crack tip stress (Wiederhorn and Bolz 1970). According to the chemical reaction rate theory (Glasstone \etal 1941), the preexponential term represents the activity of the reactants in the moist atmosphere and the exponent $m$ (generally close to 1) is related to the nearly first order of the stress-corrosion reaction (Wiederhorn \etal 1974).

For propagation in liquid environment, the term $p/p_0$ is substituted by the activity of water molecules which is strongly dependent on the composition of the liquid environment. Notably, it is strongly affected by the pH and ionic concentration of some species related to the glass composition (Wiederhorn and Johnson 1973).
Apart from water, other molecules that are effective for stress-corrosion such as ammonia, hydrazine and formamide have in common (a) the capability of donating both a proton and an electron at the two ends of a molecule and (b) a molecular diameter inferior to 0.5 nm (Michalske and Bunker 1987). The first property is necessary to present a concerted reaction of adsorption and scission of the siloxane bridges of the glass network (Fig.\ \ref{fig:SCSReaction}, bottom). The small diameter is necessary for the reactive molecules to be able to reach the strained bonds at the crack tip without steric problems. The effect of other non reactive liquids is mainly that of reducing the effective concentration of water to traces and thus of substantially suppressing the stress-corrosion propagation (Freiman 1975) to lead to a propagation similar to that in dry nitrogen or in vacuum.

\subsection{Region II: transport limited propagation}
\label{sec:RegionII}

In region II the crack propagation velocity is substantially independent of the stress intensity factor (Fig.\ \ref{fig:KVHGlass}), since the reaction rate is limited by the transport kinetics of reactive molecules towards the crack tip. The plateau value for propagation in moist air is increased when increasing humidity %or when decreasing the absolute pressure,
in qualitative accord with the proposed equation (Wiederhorn 1967):

\begin{equation}
\label{eq:RegionIIWied}
v^{II} = v_0 {p_{H_2O} D_{H_2O}} %  \over m
\end{equation}
where $D_{H_2O}$ is the diffusion coefficient of water molecules in air. However, a finer analysis reveals that several mechanisms can act in limiting the transport of the reactive molecules as a function of the environment (total pressure, humidity, temperature, liquid or vapour form of the water molecules) and of the progressive confinement experienced when approaching the crack tip (Lawn 1993). The diffusion of water molecules in a vapour or solute will be first reduced by the decrease of the average free path, and then possibly by the activated adsorption on the crack walls when the confinement becomes of the same order as the molecular size. While the first mechanism is generally accepted for glasses (Michalske and Bunker 1987), the second one is likely to be determinant in the cleavage of mica under humidity (Wan, Aimard \etal 1990).

Region II is very limited for samples plunged in liquid water, and the extension of the region II in glass was shown to be reduced at elevated humidity (Wan, Lathabai \etal 1990) suggesting the presence of capillary condensation in the crack cavity. In liquid solutions, the position of region II depends on the water concentration, and a region II plateau remains generally visible in a very low velocity region for inert liquid or gas environments due to trace amounts of water as discussed above (Freiman 1975).
Water molecules may also penetrate the strained glass network in a molecular form (Doremus 2002) and react with the network in a more internal location near the crack tip (Marsh 1964b; Tomozawa 1996). In this case the bulk diffusion rate could also act as a limiting factor in region II. However, the kinetics of most of these phenomena is ruled by the thermal activation of reactions with a stress dependent energy barrier, and is thus  expected to depend on temperature and stress in a similar manner (cf.\ section \ref{sec:chemistry}), making the discrimination difficult if $v(K)$ curves are only measured.

\subsection{Region III: inert propagation and failure}
\label{sec:RegionIII}

When reaching region III, the crack propagation velocity starts increasing again with $K$ (Fig.\ \ref{fig:KVHGlass}). The dependence is so steep that the measurements require special techniques, such as high speed cameras or a modulation of the fracture surface generated by sound waves (Gehrke \etal 1991). The position of region III is substantially independent of the local environment (except for a dependence on the dielectric constant as shown by Wiederhorn \etal 1982), in agreement with the expected suppression of the stress-corrosion reactions.
The propagation in vacuum or in inert liquid or gas environment extends the region III domain down to very low velocities (Wiederhorn \etal 1974). These delicate measurements showed that the $v(K)$ curves are still temperature dependent and that they can be fitted with an Arrhenius relation with a more elevated activation volume.

Although the velocity is rapidly climbing to large values, the propagation is still of a subcritical type, i.e.\ the time scale for crack propagation is still determined by the time required for thermal fluctuations to overcome the weaker and weaker energy barriers.
According to lattice trapping models (Thomson \etal 1971),
the critical stress intensity factor $K_c$ represents the end of the region III, i.e.\ the moment where no more energy barriers exist and the mechanical energy is converted into kinetic energy in an unstable manner, the time scale for propagation being now determined by dynamic equations. However, due to the elevated slope of region III in glasses and to the high values of the crack velocity, it is hardly possible to establish experimentally the end point of region III and the value of $K_c$ is generally identified with the position of region III itself.
From a practical point of view, when loading a sample in tension, the sample will generally fail abruptly when approaching $K_c$, and the measured value of $K$ at failure is called the fracture toughness. On the other hand, in stable fracture configurations ($dK/dc<0$, $c$ being the crack length) such under DCDC conditions, and in the absence of stress-corrosion (for example in vacuum or in nitrogen) the crack will simply arrest for $K =K_c$ and further propagate only upon an increase of the load.

The toughness of silicate glasses is generally around 0.8 MPam$^{1/2}$, within a range of $\pm 20\%$ for most typical glass compositions (Vernaz 1978, 1980)\footnote{
We should remark that the corresponding fracture energy of glass in inert atmosphere is of the order of $G_c = 2\gamma_f = K_c^2/E' \simeq 8$ J/m$^2$, which is ten times higher than the typical values of the surface tension of glasses $\gamma_0 \simeq 0.5$ J/m$^2$ (cf.\ sect.\ \ref{sec:plasticity}).}.
The relation between $K_c$ and glass composition is quite complex since the glass network structure is strongly affected by the percent concentration of each oxide. Since the toughness is generally related to the Young's modulus, according to the estimation $K_c = \sqrt{\gamma E}$, the compositions with higher Young's modulus are generally tougher. Another apparently consistent correlation appears between the decrease of toughness and the number of non-bridging oxygen atoms, which is determined by the concentration of modifier oxides (Rizkalla \etal 1992). Moreover, the toughness tends to increase when the size of the alkali ion in ternary glasses decreases (Gehrke \etal 1991), suggesting the importance of ion migration even under the rapid fracture (cf.\ section \ref{sec:IonMigration}). However, a clear comprehension is still lacking, and several research efforts are currently underway to establish clearer correlations (Kingston and Hand 2000; Kurkjian \etal 2003). The major reason for this delay is that the measurements of $K_c$ are quite delicate and that they are generally obtained with very different techniques, the overall scattering being comparable to the strength of the observed concentration effects.

\subsection{Region 0: threshold behaviour, healing and aging}
\label{sec:threshold}

\begin{figure}[!h]
\centering
\begin{minipage}[t]{0.48\linewidth}
   \centering
   \includegraphics[width=7 cm]{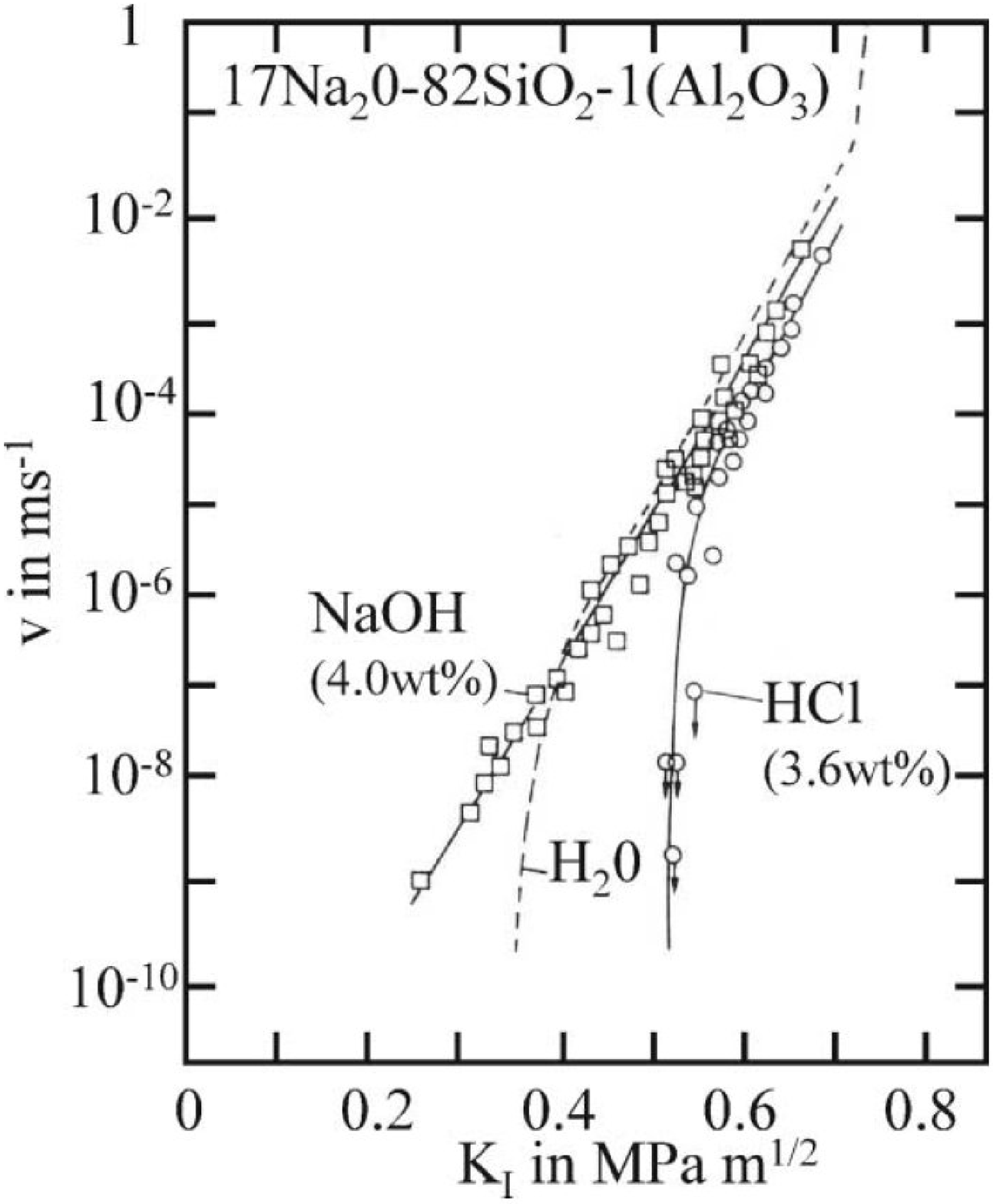}
   \caption{Threshold behaviour in a sodium-alumino-silicate glass in water and in solutions with different pH (from Gehrke \etal 1991).}
   \label{fig:Threshold1}
\end{minipage}%
\hfill
\begin{minipage}[t]{0.48\linewidth}
   \centering
   \includegraphics[height=7 cm]{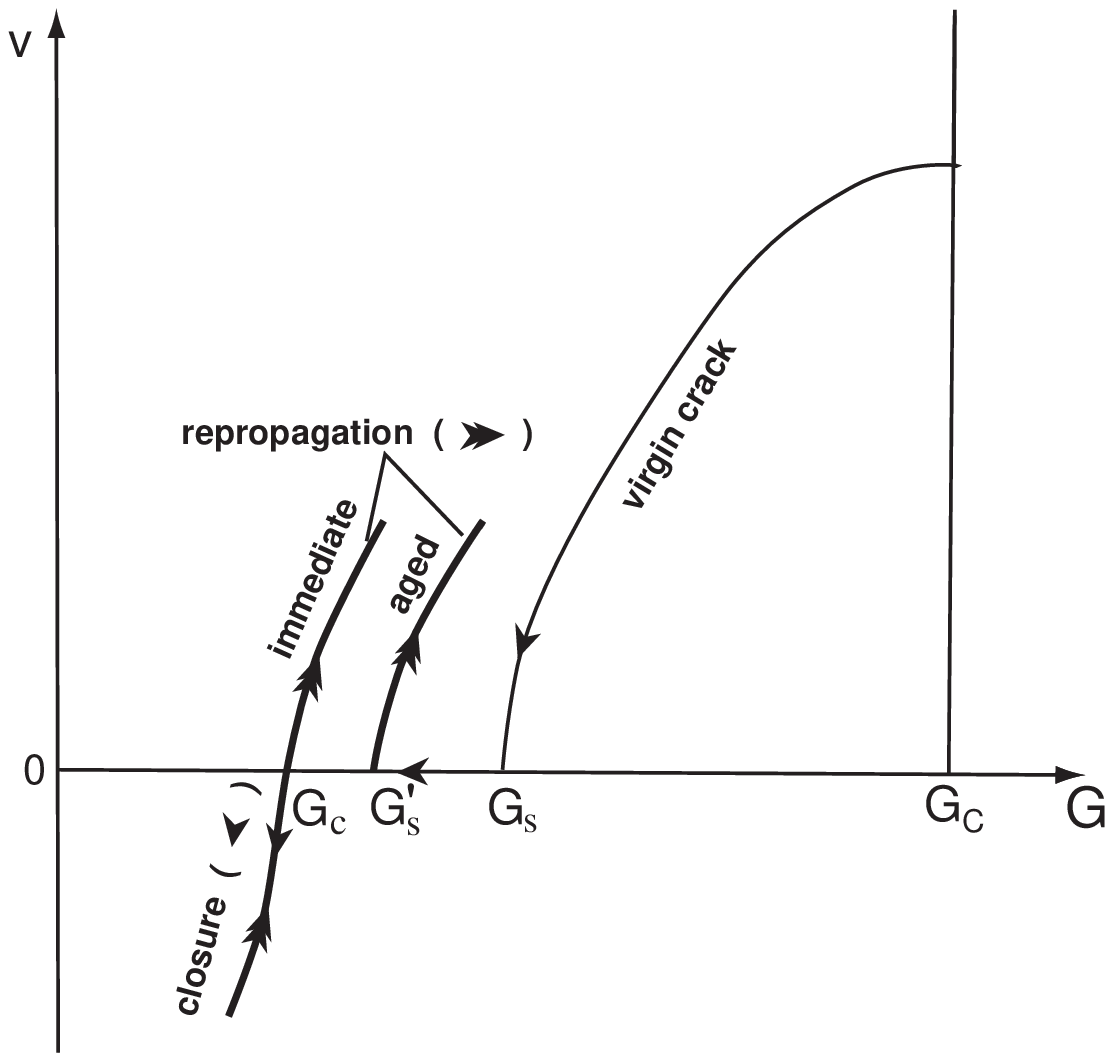}
   \caption{Sketch of the hysteretic healing and repropagation behaviour in soda-lime glass.}
   \label{fig:Healing}
\end{minipage}
\end{figure}

In some glasses, such as alkali rich glasses, a threshold behaviour occurs (Fig.\ \ref{fig:Threshold1}), i.e.\ fracture velocity rapidly falls to undetectably low values
when $K$ decreases toward a threshold value $K_s$ (Gehrke \etal 1991). This property is of extreme importance in the life duration of engineering glass products, yet its origin is very subtle and has long been subject to debate.

According to Griffith theory, a threshold for propagation should exist at $G = G_0$ where $G_0=2 \gamma_0$ represents the thermodynamic surface energy which is necessary to create the new surfaces through the breaking of the cohesive bonds. In the framework of the rate reaction theory (Glasstone \etal 1941) this equilibrium condition corresponds to the equality of the rates of the two opposing reactions of breaking and recombination. The crack propagation velocity in proximity to such an equilibrium threshold can thus be approximated by including in Eq.\ (\ref{eq:PolletBruns}) a term for the rate of recombination (Lawn, 1993):

\begin{equation}
\label{eq:GrifG0}
v = v_0 \exp \left( - { \Delta E_a - \alpha (G - G_0) \over k_BT} \right) + v_0 \exp \left( - { \Delta E_a + \alpha (G - G_0) \over k_BT} \right)
\end{equation}

$$=  v_0 \exp \left( - { \Delta E_a \over k_BT} \right) \sinh \left( { \alpha (G - G_0) \over k_BT} \right)$$
Equation (\ref{eq:GrifG0}), besides presenting a threshold for propagation at $G=G_0$, also implies the backward propagation or healing of the crack for $G<G_0$, in agreement with the thermodynamic considerations of Rice (1978)\footnote{
We should remark that we can only talk about reversible crack propagation for $G\approx G_0$, i.e.\ for vanishing propagation velocity, while both forward and backward propagation with finite velocity imply a rate of energy dissipation $(G-G_0)v \geq 0$.}.

However, this scenario is not exactly what is happening with glass. For soda-lime glass, although the crack propagation stops at a threshold $G_s \simeq 0.8$ J/m$^2$, no healing is observed down to a lower value $G_c\simeq 0.15$ J/m$^2$ (cf.\ Fig.\ \ref{fig:Healing}), suggesting that the crack closing occurs due to a different mechanism from crack opening, for instance through the hydrogen bond interaction between hydrolysed crack surfaces (Stavrinidis and Holloway 1983). For silica glass, although no propagation threshold $G_s$ is observed down to $v = 10^{-13}$ m/s (Muraoka and Abé 1996), crack closure also happens for a value of $G_c$ similar to that of soda-lime glass.
Moreover, the repropagation kinetics is quite different in the two kinds of glasses. While repropagation of the healed crack in silica happens as soon as $G \geq G_c$, the repropagation threshold $G'_s$ in soda-lime progressively increases between $G_c$ and $G_s$ as a function of the time spent in the closed configuration (Stavrinidis and Holloway 1983), suggesting partial reformation of the siloxane bridges (Michalske and Fuller 1985).

This phenomenon should not be confused with another kind of aging that can be observed in soda-lime glass. When the crack is arrested and held for a given time at a value $G_h$ in the interval $G_c<G_h<G_s$ (NB: without healing), the repropagation will not start again at $G_s$, but it will either need the application of a higher value of $G$, or it will occur after a delay time which is maximum when $G_h\simeq G_s$ (Michalske 1977, Gehrke \etal 1991).

The threshold behaviour and the delayed fracture after crack arrest in alkali containing glasses were shown to be fundamentally related, and can be interpreted as a consequence of the ion exchange phenomenon (Gehrke \etal 1991; Fett \etal 2005a). Ion exchange at the crack tip and at crack surfaces can have several effects: (a) the build up of compressive stresses at the crack tip, shielding the external stresses, (b) a change of the surface energy and/or a dissipation of energy that can alter the energetic balance, (c) a change of the local glass composition leading to an increase of the fracture resistance.
The ion exchange interpretation was supported by the strong dependence of the threshold and arrest time for soda-lime in liquids on the pH of the solution and the glass composition in a series of binary and ternary glasses (Fig.\ \ref{fig:Threshold1}). Thus, an acidic solution ($HCl$) enhances the ion exchange and causes the threshold to shift to higher values of $G_s$, while a basic solution ($NaOH$) can make the threshold drop below the detection limit. A specific exception is observed in glasses with high alkali content, where the corrosion behaviour is so rapid that fracture propagates spontaneously, presenting a horizontal plateau on the $G(v)$ curve for a low limit velocity around $10^{-7}$ m/s, especially in presence of an acidic solution (Gehrke \etal 1991).

The phenomenon of ion exchange will be discussed in more detail in sections \ref{sec:chemistry} and \ref{sec:IonMigration}, firstly because the effect of local crack tip stresses plays an important role, and secondly because the discussion of the difference of behaviour between the propagation in water and in moist atmosphere requires a more detailed analysis of the chemical processes in confined liquid films. However, some considerations are opportune here. The hysteretic behaviour of the crack closure and repropagation in soda-lime glass,
as well as the non observation of any threshold in silica, cast strong doubts on the applicability of the Griffith scenario for reversible healing in glasses.
The hydrolysis reaction of the siloxane bonds is essentially irreversible at ambient temperature, and the crack closure at $G_c$ principally involves the adhesion of the hydrated crack surfaces (the latter being fairly reversible for silica). For alkaline glasses, the modifications of the crack surfaces are even more drastic due to the irreversible production of ion exchange layers. It is therefore not easy to associate the position of the threshold $G_s$ with a glass surface energy in the Griffith sense, but its value appears to arise from a competition between the kinetics of the formation of the ion exchange layer and the crack propagation kinetics. It is however difficult to disregard the similarity between the typical values of $G_s\sim 0.8$ J/m$^2$ and the estimates of the glass surface tension, both being about one tenth of the fracture energy in inert environment. This seems to suggest that the complex interaction of glass with water during slow fracture helps the glass surface to relax from the elevated non equilibrium energies towards an equilibrated condition.

\section{A deeper insight into the stress-corrosion and damage mechanisms}
\label{sec:Insight}

\subsection{Competition between dissolution, corrosion and leaching}
\label{sec:chemistry}

In order to understand more fully the different components of the chemical contribution of water molecules to the corrosion and damage of glass, the three main categories of reactions that can occur between glass and an aqueous solution should be reexamined (cf.\ Bunker 1994 for a more detailed review): (1) hydration consists of the penetration of molecular water into the glass network as an intact solvent; (2) hydrolysis and condensation are the two opposite chemical reactions which break or reform the $Si-O-Si$ (or equivalent) network bonds by exchange with a couple of hydroxyl groups according to the scheme:

$$Si-O-Si + H_2O \Leftrightarrow Si-OH + HO-Si$$
(3) ion-exchange reactions consist of the replacement of a modifier cation (such as sodium) by a hydrogen or hydronium ion:

$$Si-O-Na^+ + H_3O^+ \Leftrightarrow Si-OH + Na^+ + H_2O$$

A good account on the solubility of silicate glasses can be found in Iler (1979). The dissolution reactions are so slow at ambient condition that glasses are considered as exceptionally inert materials, and generally need tens of years exposure to moisture before showing appreciable surface alteration.
However, when increasing temperature or stress these reactions can be significantly accelerated, especially under extremely acid or basic environmental conditions. Moreover, if the three reactions occur at the same time, each reaction will influence the kinetics and mechanisms of the other reactions (Bunker 1994).

In pure silica glass only the first two reactions are possible and they cooperate in the water diffusion kinetics. According to the reactive diffusion model developed by Doremus (well reported in his book of 2002), water mainly diffuses in molecular form with a well defined thermally enhanced diffusion coefficient, but it continuously reacts with the glass network being thus temporarily immobilised in the $SiOH$ form. As a result, an effective water diffusion coefficient can be defined that also depends on concentration. At elevated temperature, the water content converges to the equilibrium solubility (mainly dependent on the free volume in the glass network) and the hydrolysis reaction is equilibrated (mainly towards the dissociated state). However, at lower temperatures close to ambient condition, the reactions are not equilibrated and both hydroxyl and molecular water are present in the glass with a very complex speciation, which is often subject of debate (Stuke \etal 2006; Doremus 2002).

In alkali containing glasses such as soda-lime, water diffusion is greatly complicated by the combination with the ion exchange reactions (Doremus 1975). Since alkali ions partly fill the network voids, ion exchange can be the limiting factor for water diffusion, but hydrolysis is also necessary to open up the pathways in the glass network (Bunker 1994). Lanford \etal (1979) have shown that alteration layers are formed on glass surfaces in liquid water by selective leaching of alkali ions. The layer width is initially proportional to the square root of time (diffusion controlled process) and eventually reaches a stationary value due to the equilibration with the slow etching of the glass surface.
The hydrated layers that are formed on soda-lime glass were shown to be sodium-depleted in a proportion that suggests the substitution of sodium ions by the hydronium ions $H_3O^+$ (rather than the hydrogen ion).
This exchange is responsible for the development of a stressed surface layer depending on the relative volume difference of the exchanged ions. In the case of soda-lime glass the developed stress is compressive (Michalske and Bunker 1989, Fett \etal 2005b), in agreement with the hypothesis of a sodium/hydronium exchange, the volume of hydronium and hydrogen being respectively larger and smaller than that of sodium. In an analogue manner the exchange of sodium with the larger potassium ions is induced in the chemical tempering treatments to enhance the strength of glass products (see Gy 2008 for an extensive review).
The composition and structure of the leached layer are thus heavily altered and are more similar to a hydrated silica glass than to the initial glass.
Moreover, under extensive leaching, the recondensation reactions can also become important actors in a more profound restructuring   of the glass network leading to the formation of a more compact glass phase, similar to anhydrous silica glass, pervaded by a more permeable porous network, often called a gel layer (Bunker 1994).

While the behaviour and thermodynamics of these corrosion processes have been extensively studied as functions of temperature, pH and composition of the liquid in contact with the glass surface (cf.\ the review of Conradt 2008), the behaviour under moist atmosphere is often more subtle (Bunker 1994) due to the formation of thin nanometric water films on glass surfaces whose detailed chemistry escapes state of the art thermodynamics and often entails the formation of complex patterns (Watanabe \etal 1994).
Moreover, although these reactions are known to be strongly influenced by the local stress state, the accurate study of their behaviour at elevated tensile stresses is still lacking and is subject of strong debate. In the next sections we will try to clarify the present knowledge about
the damage processes near the crack tips, where the extremely elevated stress gradients affect smaller and smaller zones, thus making the space and time scales of the different mechanisms be highly variable and interdependent.

\subsection{Crack tip blunting}

The initial sharp-crack atomic-bonding paradigm really means that at room temperature the hydrolysis of individual crack-tip bonds, which is singularly enhanced by the elevated crack tip stress, is the only relevant process for determining the crack tip propagation laws (cf.\ Fig.\ \ref{fig:SCSReaction}). Alternatively, this condition at least should hold once we have identified the stress intensity factor $K^*$ in the local enclave, thus admitting the possibility that the other processes and reactions are effective in a surrounding process zone (Fig.\ \ref{fig:MechSchemeA}).
This hypothesis constitutes a singular (degenerate) version of the original Charles and Hilligs (1962) theory, in which the stress-enhanced corrosion rate can not act any further in sharpening the crack tip since its curvature radius has reached a lower limit, estimated to 0.5 nm and corresponding to the radius of the basic siloxane rings of the glass network.

The assumed condition of propagation at constant tip radius is the basis for the interpretation of the Wiederhorn equation (\ref{eq:Wiederhorn}) in terms of the LEFM theory and for the uniqueness of the $v(K^*)$ curve (Lawn, 1983). However, it can not apply to the description of the initial development of blunt flaws into sharp cracks, and we can rise the question of whether this evolution would be the only possible destiny for  progressively sharpening cracks under all conditions. Can the final crack tip radius of curvature be a function of $K$ at least for certain glass compositions and/or environments? Can the competition between the rate of the different processes and the crack tip velocity make the crack tip so blunt (or `effectively' blunt due to material damage) that it may stop under an apparently elevated value of $K$ (i.e.\ larger than that expected from the glass surface energy)?
The debate on this subject has been especially active concerning the origin of the propagation threshold and of the crack closure/aging/repropagation behaviour (plasticity issues will be discussed in section \ref{sec:plasticity}).

\begin{figure}[!h]
\centering
\includegraphics[width=6 cm]{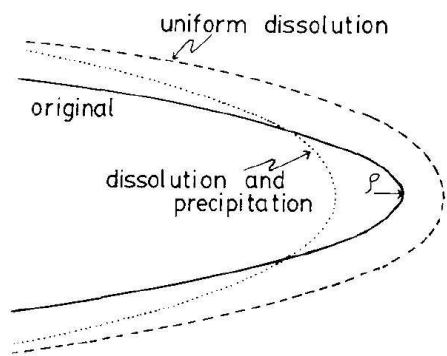}
\caption{Schematic representation of the blunting mechanisms (from Ito and Tomozawa 1982).}
\label{fig:Blunting}
\end{figure}

Since the solubility of glass surfaces is also dependent on the local radius of curvature (Iler 1979), when the external stresses are very low the crack walls should be corroded more rapidly than the tip, leading to a progressive tip blunting effect (Ito and Tomozawa 1982). Moreover, the presence of a gradient of solubility in a very confined environment can lead to a concomitant phenomenon of dissolution from the lateral walls and reprecipitation at the crack tip which would even enhance the extent of the tip blunting (cf.\ Fig.\ \ref{fig:Blunting}), as observed in a controversial TEM study of the crack tip in silica (Bando \etal 1984).
If such blunting happens under almost static conditions, then the fracture should be substantially reinitiated after reloading, leading to an effect of temporary increase of the glass strength, which was proposed as an alternative explanation to the occurrence of a time delay before repropagation of aged fractures (Han and Tomozawa 1989).

%W. D. Kingery, H. K. Brown, and D. R. Uhlmann; pp. 425-30 in Introduction to Ceramics. Wiley & Sons, New York, 1976.

The difficulty in this debate arises from different causes. On one hand, since all these reactions are influenced by stress, temperature and pH, the extent and rapidity of occurrence of these phenomena is strongly dependent on the specific conditions and on both the glass composition and environment, making the comparison of different experiments more subtle, especially if we take into consideration the difference between the behaviour in water and in moist atmosphere.
On the other hand, several conclusions were based on the indirect effect of these phenomena on the measurements of strength of glass samples subjected to different kinds of treatments (Han and Tomozawa 1989). A condition which has often been shown to lead to several superposed effects on the flaw distribution and nature in the glass, but also on the local state of the residual stresses which strongly affects these types of measurements. For example, in the case of the aging of indentation flaws, the experiments of Lawn \etal (1985) suggest that the strength increase during aging is caused by the relaxation of the indentation residual strains (also produced to a variable extent when finishing the glass surface) through subcritical crack propagation of the flaws.

Fett \etal (2005b) have recently modelled quantitatively the strength increase of an uncharged soda-lime glass plate after inducing a surface compression layer by ion exchange in water at 90 C, and obtained good agreement with experimental measurements. They also made an analogue model for the effects of the formation of an ion exchange layer in the internal crack walls and tip neighbourhood of both static and propagating cracks in a soda-lime glass in water (Fett \etal 2005a). This model supports the interpretation of the threshold behaviour by the effect of the compressive stresses induced in the ion exchanged layer. They finally compared (by AFM {\it post-mortem} recombination) the mismatch in the morphology of the crack surfaces obtained after inducing the partial corrosion of the crack walls of an arrested crack in soda-lime glass (Guin and Wiederhorn 2005) and letting the crack repropagate to failure. The comparison reveals a homogeneous corrosion of the two faces and is argued to be evidence for the sharp nature of the crack tip at the propagation threshold.

Although these measurements are quite convincing for the specific case of soda-lime glass, the applicability to all conditions requires more systematic work of this kind. It is generally accepted that after extreme corrosive attack (such as after HF etching) the cracks are very likely to be blunted as proposed by Proctor (1962). However, several intermediate conditions may exist where the elevated susceptibility of some kinds of glass to dissolution induces an anomalous propagation behaviour as in the high alkali binary glasses studied by Gehrke \etal (1991) or in some alumino-phosphate glasses (Etter \etal 2008).

\subsection{Crack tip plasticity}
\label{sec:plasticity}

Another major objection to the sharp-crack atomic-bonding paradigm comes from the hypothesis of a significant contribution of plastic deformation at the crack tip in the slow crack propagation mechanisms.
Since the early works of Dugdale (1960) on metals, and the observations of plastic behaviour of glass under compression in indentation and scratch marks (Taylor 1949), several investigations have been carried out to track the evidence of plastic behaviour in the strong tensile stress field at crack tips in glasses. In particular, Marsh (1964a; 1964b and refs there in) edited a series of papers developing a complete theory of elasto-plastic fracture propagation for glass, which was proposed as an alternative to the Griffith explanation for brittle glass strength and to the Charles-Hilligs interpretation of static fatigue as the effect of stress-corrosion.

The theory closely follows the formalism developed by Dugdale for metals, defining as a suitable fracture criterion the formation of a plastic zone of critical dimensions $R \simeq 6$ nm (Marsh 1964b) for typical glasses having an estimated yield stress of about 6 GPa. The occurrence of static fatigue is explained as a contribution of the stress-enhanced water diffusion into the glass to a local reduction of the glass viscosity and thus of the resistance to plastic flow (Marsh 1964b). The time scale for fracture propagation is thus attributed to the transport of water in bulk glass and to the consequent viscous flow, although the kinetics is not explicitly developed.

Wiederhorn (1969) performed accurate measurements of the fracture energy for six glasses in inert environment and showed that once the stress-corrosion is suppressed, the fracture surface energies $\gamma_f=G_c/2$ range between 3.5 and 5.3 J/m$^2$, a figure
which is ten fold higher than the typical values of the surface tension of glasses $\gamma_0 \simeq 0.5$ J/m$^2$ estimated by Griffith (1921) and which suggests a major contribution of irreversible processes in crack propagation. By analysing the implications of the plastic model proposed by Marsh (1964b), Wiederhorn (1969) estimated a plastic work of 4.5 J/m$^2$ for a plastic zone of 2.6 nm and a yield stress of 10 GPa, and he observed that although it is in rough agreement with the measured fracture energies it is very small compared to that obtained in metals of $10^4$ J/m$^2$. Moreover, since all the estimated space scales for the local deformations are close to the dimensions of the molecular scale (the average size of glass network rings is of the order of 0.5 nm), he questioned the opportunity of applying the concept of `plasticity' to the nanoscale.

On the other hand, the sharp-crack atomic-bonding paradigm that was advocated by Lawn (1983) as the foundation of the brittle fracture theory, considers that load is borne in a completely elastic way at the molecular level. According to this hypothesis only the molecules at the direct end of the molecularly sharp crack tip are broken sequentially under the elevated load. Lawn \etal (1980) conducted an extensive evaluation of this possibility by comparing the nature of the atomic bonds and structure between glasses and other brittle or ductile materials.
They supported their conclusion on the absence of plastic behaviour in glasses demonstrating by TEM observations the absence of dislocations in crystalline oxides with same composition (as opposed to more ductile crystals).

However, the hypothesis of Marsh (1964b) went well beyond the discussion of fracture in inert atmosphere and vacuum conditions.
Marsh was well aware that water is the oxide which has the highest effect in lowering the viscosity of a glass melt (cf.\ Del Gaudio \etal 2007) and his explanation of the effect of humidity on the slow crack propagation (as an effect of reduced glass viscosity by stress-enhanced diffusion of water into the glass) was intended as a completely alternative mechanism for stress-corrosion. By looking at this interpretation with more modern eyes, the basic mechanism remains indeed similar, since the reduction of glass viscosity is probably caused by a stress-enhanced hydrolysis of the network siloxane bridges, substituted with more mobile $SiOH$ terminations. However, this hydrolysis would not be limited to the crack tip rings, but rather involve a larger process zone, of the order of a few tens of nanometres, corresponding to a region with an effective yield stress of 1 GPa. The model is also quite different in that the time scale is not simply governed by the rate of reaction at the tip, but also by the rate of stress-enhanced diffusion of water into the glass and the time scale of structural relaxation which is accelerated by both stress and the reduced viscosity due to hydration. This would lead to a process zone size that is a decreasing function of the crack velocity and thus of $K$.

The development and experimental investigation of a similar scenario was progressively carried on by Tomozawa and collaborators in the eighties using IR and Raman spectroscopies to detect water diffusion profiles in silica glass and the consequent changes in the glass structure, identified by changes of the so called `fictive temperature'%
\footnote{Glass being an out of equilibrium material, its structure and properties are not uniquely determined by the thermodynamic variables, but also depend on the past thermal history. According to the hypothesis of Tool (1946), the thermal history may be represented by the knowledge of a unique additional variable named the fictive temperature $T_f$. This variable represents the temperature of the supercooled equilibrated glass melt which has the same structure as the present glass. For the same composition, a higher fictive temperature corresponds in general to a lower density and index of refraction. When a glass is aged at a temperature $T'<T_f$, its structure undergoes a slow evolution toward the equilibrated state, the relaxation time being inversely proportional to the glass viscosity. During this relaxation the fictive temperature decreases towards the value of $T'$. However, in practice it is quite difficult to equilibrate glasses to a fictive temperature lower than 900 $^\circ$C.}. A good review on this work can be found in Tomozawa (1996). However, some relevant points will be recalled here.
Ito and Tomozawa (1981) initially observed an increase in the dissolution rate of silica glass under hydrostatic pressure at 285$^\circ$C, and argued that the opposite should happen in tension, in contrast to the stress-corrosion hypothesis of Charles and Hilligs (1962). On the other hand, Nogami and Tomozawa (1984) showed that the water diffusion into glass was exponentially enhanced on the tensile side of a bent plate of silica glass under saturated moist atmosphere at 192 $^\circ$C (the opposite occurring on the compressive side), the hydrated layer reaching a thickness of several microns after 300 h. These two combined observations led Tomozawa to suggest that glass structural relaxation caused by stress-induced water penetration at the crack tip could be an alternative mechanism for slow crack propagation in glass. However, most measurements were done at high temperature and the extrapolation to ambient condition is generally difficult due to non equilibrated reactive diffusion and to difficulties in estimating the solubility, especially due to reported trend inversions at different temperatures.
A major advance was provided by Tomozawa \etal (1991) who measured by resonant nuclear reaction analysis the enhanced water entry into silica glass fracture surfaces during slow crack growth at ambient temperature. The thickness of the hydrated layer at ambient condition was estimated at about 10 nm which is of the order of the experimental resolution, but the increase in the amount of water is significant when compared to crack propagation in inert environment.
The support for the overall model has progressively advanced through several complementary observations, which are necessary to separate and test the effect of the water content, the fictive temperature and the applied stress. Peng \etal (1997) have documented the effect of stress and water content in accelerating the surface structural relaxation in silica fibres. Li \etal (1995) have shown that silicate glasses with higher fictive temperature present both greater fatigue resistance and greater inert strength. Koike and Tomozawa (2006) showed that the fictive temperature also affects differently the subcritical crack growth rate of silicate glasses and soda-lime glasses.

The bibliography on this subject is very rich and can not be reviewed in detail here. Several observations and especially some interpretations have been long debated and are still disregarded by the `sharp-tip' community, since once again they are based on many indirect observations and extrapolations from different temperatures and other conditions. Moreover, the thermal treatments to change the fictive temperature are not without effects on the flaw distributions and the effects on strength should be interpreted with great care. However, the Tomozawa mechanism remains largely plausible and would need a major synthetic effort to prove its real effectiveness for crack propagation in silica (and especially in other glasses) at ambient condition. Such modelling should end in the writing of kinetic equations that would fit the experimental data as well as the Wiederhorn equation (\ref{eq:Wiederhorn}) and that would provide a physical interpretation for the model coefficients related to measurable quantities.

A different series of investigations were made possible by the application AFM techniques in glass science in the '90s.
In 1996 Guilloteau \etal made the first {\it in-situ} AFM measurements of the external glass surface during crack propagation in borosilicate glass. They observed the presence of a surface depression ahead of the crack tip, and they estimated the presence of a process zone of 50 nm size (corresponding to a yield stress of 1 GPa) by identifying the deviation of the vertical displacement profiles from the power law predictions of linear elasticity. Similar results were obtained by Célarié \etal (2003) on lithium alumino-silicate glasses and by Prades \etal (2005) on silica glass, by {\it in-situ} AFM observation of slow fracture propagation in DCDC samples under pure mode I. The process zone size was observed to grow in size from 20 to 100 nm when the propagation velocity diminished from $10^{-10}$ to $10^{-12}$ m/s due to a decrease of $K$. Moreover, these AFM measurements suggested that the crack propagation in the process zone proceeds by the nucleation, growth and coalescence of nanometric cavities, in a similar way to what happens in the ductile fracture of metals at the micrometer scale.

These observations reopened the debate on crack tip plasticity, which is at present very active due to several contradictory observations.
Lopez-Cèpero \etal (2007) have observed that the {\it post-mortem} recombination of the AFM measurements of the morphology of opposing crack surfaces in silica and soda-lime showed no evidence for the expected traces of the nanocavities in the bulk of the specimen.
Moreover, Fett \etal (2008) showed the inadequacy of the 2D plane-stress solution of Maugis (1992) used by Guilloteau \etal (1996) and then by Célarié \etal (2003) to represent the elastic vertical surface displacement near the crack tip with a $r^{-1/2}$ dependence.
When using the correct $r^{\lambda}$ dependence (with $\lambda \simeq 1/2$), which was derived by Bazant and Essentoro (1979) to describe the corner singularity of a 3D surface striking crack (cf.\ Dimitrov \etal 2006), the deviation of the displacement profiles measured by AFM from the elastic solution is much less evident (Fett \etal 2008). The quality of the fit on the vertical topography profiles is deeply affected by the presence of the residual roughness even after excellent polishing. Moreover, the combination of the roughness and the crack tip surface deformation could produce very similar features to the cavities observed by Célarié \etal 2003 as shown by numerical simulation (Fett \etal 2008).

However, all these measurements are very close to the instrumental limits of the AFM topographical images which are made with AFM tips of an average radius of 10 nm. Better insights could in principle be obtained through the delicate approach of separating the vertical surface displacement from the surface roughness by using digital image correlation techniques (Hild and Roux 2006). Moreover, a signature of the occurrence of inelastic processes during crack propagation can be identified through the statistical correlation functions of the crack surface morphology (cf.\ section \ref{sec:surfaces}).

The recent development of complementary techniques to measure the crack tip stress fields with nanometric resolution are also very promising for determining the effective scale of a damage process zone. We can cite the SEM cathodo-luminescence measurements (Zhu \etal 2007; Pezzotti \etal 2008) and the promising measurements by nanoRaman and electron back scattered diffraction techniques (Vaudin \etal 2008).

\subsection{Local crack tip environment}
\label{sec:tipenvironment}

Since crack propagation in the stress-corrosion regime is strongly affected by the environment at the crack tip, and since this is a highly confined region with a typical crack opening ranging from 0.5 nm at the tip to a few nm at a distance of 1 $\mu$m, it is questionable whether the local crack tip environment can be treated as equivalent to the outer environment, be it liquid or gas, and also whether the macroscopic description of the thermodynamic state of the environment at a nanometric confined crack tip is relevant.

The region II behaviour is a first sign of an inhomogeneous crack tip environment. Since water is being used by the crack tip for propagation, the local concentration (in gas or solute) at the crack tip is considered to decrease when the crack tip velocity is comparable with the transport rate of water in the cavity. The near tip concentration in region II was modelled to fall to near zero (Wiederhorn 1967), and the concentration gradient established in the crack cavity would be the motor for water diffusion to the crack tip, its rate becoming the controlling factor for crack propagation (cf.\ section \ref{sec:RegionII}). On the other hand, when the whole sample is plunged into water, a more subtle issue is given by the viscous transport of liquid water toward the tip. This transport is indeed very effective since the region I then only starts changing its slope when near to region III. However, at elevated velocity, a pressure drop at the tip may be caused by the increasing viscous drag, leading eventually to a negative pressure in the crack tip region. A very interesting investigation by Michalske and Frechette (1980) has suggested that a strongly negative pressure can be sustained at the crack tip without cavitation due to the extreme confinement, leading to the observation of a supercritical crack propagation at reduced velocity, which can be interpreted according to the shielding effect (reduction of $K^*$) caused by the internal forces in the crack cavity (cf.\ Fig.\ \ref{fig:MechSchemeA}). A further increase of $K$ eventually leads to cavitation in water when the negative pressure extends to a less confined region of the crack cavity, followed by an instantaneous order of magnitude rise in the crack velocity due to the sudden depletion of the crack tip environment and to the release of the internal forces. A similar cavitation-induced dynamical instability was observed by Maugis (1985) in brittle polymers.

In alkali containing glasses in water, the enhanced leaching at the crack tip may cause a local change of the liquid composition at the crack tip. This was suggested to be the cause of an observed transition in the slope of the region I in soda-lime glass (Wiederhorn and Johnson 1973). At low crack velocity the crack tip environment would be equilibrated to the external liquid composition, while becoming more related to the crack tip reactions at higher velocity. The same kind of effect can be supposed for other type of corrosion products, such as silicic acid, but still has to be quantified.

Recent AFM {\it in-situ} observations have shown that nanometre scale capillary condensation occurs at crack tips in silica glass in moist air for crack velocities between $10^{-12}$ and $10^{-9}$ m/s (Wondraczek \etal 2006; Ciccotti \etal 2008). This liquid phase is made stable by the short range interaction with the glass surfaces under elevated confinement and constitutes a major alteration of the local environmental condition at the crack tip with respect to the gaseous hypothesis (Wiederhorn 1967). In this condition, not only is the water readily available at the crack tip, but the local pressure becomes strongly negative due to the Laplace pressure, thus exerting a strong attractive force between the crack lips and altering the equilibrium of the stress-corrosion reactions. Since the crack-tip reaction rate is expected to depend on the chemical activity of water, and the condensate was shown to be in equilibrium with the humid atmosphere (Grimaldi \etal 2008), we can expect that the humidity dependence of the crack velocity remains similar to what expected in gaseous environment, yet the situation should change when transport phenomena become relevant or when the equilibrium is questioned. For example, the liquid condensate can be an explanation of the reduction of the transport limited region II region observed at elevated humidity (Wan, Aimard \etal 1990). The limiting factor for the diffusion of water is shifted from the very confined crack tip region to the end of the liquid condensate where the larger crack opening allows faster diffusion.

The liquid condensation was also observed on alkali containing glasses (Célarié \etal 2007) and should in this case have a much larger effect since the local compositional changes, due to the crack tip enhanced leaching, can not be diluted like in a bulk liquid. Preliminary results show that the extent of the condensation in soda-lime glass is significantly enhanced by the alterations due to the local leaching (cf.\ section \ref{sec:IonMigration}).
The relevance of capillary condensation in the propagation kinetics of phosphate glasses, which are less resistant against dissolution, can be seen in the anomalous temperature behaviour observed in the $v(K)$ relations measured by Crichton \etal (1999) and similarly by Etter \etal (2008) at ambient temperature. The anomaly is suppressed at higher temperature where the condensation can not form.

\subsection{Alkali ion migration under stress gradient}
\label{sec:IonMigration}

In section \ref{sec:chemistry} we have seen that chemically driven sodium ion migration can happen spontaneously at soda-lime glass surfaces thus inducing a local state of compression in a surface hydrated layer. Gorsky (1935) first showed that the opposite can also be true, i.e.\ that a strong stress gradient can induce sodium migration towards the tensile direction of the gradient. While the spontaneous slow flow is generally balanced by interchange with other positive ions (such as the hydronium ion) in order to preserve charge neutrality (Doremus 1975; Lanford \etal 1979), the behaviour under a strong stress gradient is less evident. Weber and Goldstein (1964) measured a transient electric current between the tension and compression sides of a bent soda-lime slide, indicating that the non-balanced flow can occur before being counteracted by the build up of an opposite electric potential difference. Moreover, Langford \etal (1991) have revealed intense sodium emission after rapidly fracturing soda-lime glass in vacuum. The delay in the emission was of the order of a few tens of milliseconds, indicating that sodium migration can be very fast in the presence of strong stress gradients and can thus also affect region III water-free fast crack  propagation. Several observations were made concerning an excess sodium concentration in the outermost surface layer of fracture surfaces in soda-lime, overlaying a sodium depleted region of larger thickness of hundreds of nanometres (Pantano, 1985). This phenomenon looks similar to the chemically driven migration, however its space and time scales are quite different and more investigation is needed concerning the interaction of this phenomenon with the external environment during and after the fracture process.

Watanabe \etal (1994) made an interesting AFM investigation of the time evolution of the fracture surfaces of a soda-lime glass broken in low vacuum at a velocity of $10^{-3}$ m/s (region III) and then aged in ambient moist atmosphere. The excess concentration of sodium on the fracture surfaces is the cause of an unusually rapid corrosive action on these surfaces, observable by the appearance of protuberances and swellings interpreted as the recondensation of corrosion products in the form of a weak sodo-silicate gel material. The evolution of such swellings under a common atmosphere finally evolves to the formation of carbonatic crystallites due to the interactions with $CO_2$ dissolved in the sodium rich wet layers.

\begin{figure}[!h]
\centering
\includegraphics[width=14 cm]{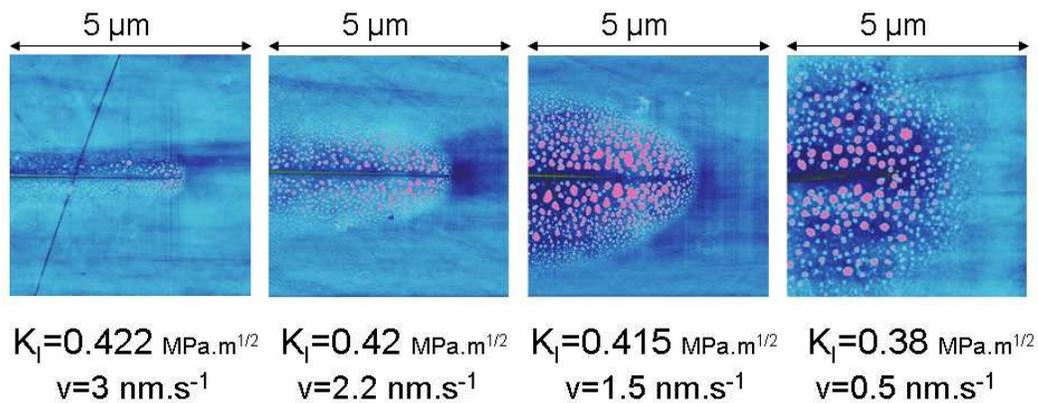}
\caption{AFM topographical images of a propagating crack in soda-lime glass at different crack velocities for RH=45\%.}
\label{fig:SodiumDiffusion}
\end{figure}

A similar kind of alteration has been observed by imaging external surfaces in the neighbourhood of a crack tip produced by indentation on a soda-lime slide (Nghiem 1998).
More recently, Célarié \etal (2007) have conducted a systematic {\it in situ} AFM study of the kinetics of the growth and evolution of the protuberances during crack propagation in soda-lime glass under controlled atmosphere of nitrogen and variable humidity levels. The absence of $CO_2$ allowed focusing on the space and time scales of the sodium exchange mechanisms and of the surface corrosion process. The region affected by the surface swellings is shown to have a parabolic shape which is modelled according to a competition between the spreading of the surface reaction and the crack propagation velocity (Fig.\ \ref{fig:SodiumDiffusion}). The parabolic shapes for different crack velocities at 45\% RH are consistent with a diffusive process with an effective diffusion coefficient of the order of 1 nm$^2$/s, which is an increasing function of the relative humidity. The application of phase imaging techniques has also allowed to identify the thickening of the sodium enriched liquid layer (predicted by Watanabe \etal 1994) prior to the manifestation of the swellings.

When fracture velocity slows down below $10^{-9}$ m/s the ion exchange front can spread ahead of the crack tip and thus start relaxing the tensional stresses in the crack tip region, which in turn will cause the progressive arrest of the crack propagation. This suggests the extension to the propagation in moist atmosphere of the interpretation of the threshold behaviour proposed by Michalske and Bunker (1985). The enhanced liquid condensation stimulated by the sodium enrichment at the fracture surfaces and in the parabolic front plays the role of the bulk liquid reservoir, but the strong confinement is likely to cause different behaviours, like in the observations of Watanabe \etal (1994). The space and time scales for the spreading of the ion exchange can be quite different.

Fett \etal (2005a) modelled the mechanical effect caused by the chemical induced ion exchange layer forming on the crack surfaces and at the crack tip of arrested and propagating cracks in soda-lime glass. This modelling is based on the ion exchange kinetics proposed by Lanford \etal 1979 for an etched soda-lime surface in water at pH7. The lateral extension of the parabolic front predicted by this model would be about 1000 times smaller than the one observed in Fig.\ \ref{fig:SodiumDiffusion}. The extremely different scales can be attributed on one side to the discussed difference of the propagation in moist air (and the consequent change in the composition of the condensed liquid layers), and on the other side to the influence of the strong stress gradient at the crack tip, which is not accounted in the model of Fett \etal (2005a).

\subsection{Fracture surfaces}
\label{sec:surfaces}

As we have seen throughout this review, {\it post-mortem} study of the fracture surfaces is an excellent complementary tool to understand the physics and chemistry of the mechanisms of crack propagation. On one side, the morphology of the fracture surface can contain relevant information on the dynamics of crack propagation and on the traces of inelastic processes or of the corrosion at the crack surfaces. On the other side, modern high resolution chemical probes can provide invaluable insights into the nature of these high energy surfaces and of the mechanisms that originated them (Pantano 1985).

The fracture surface morphology has long been used to draw information about the origin and propagation of the cracks that led a glass specimen to failure. From an optical inspection, one can distinguish three typical regions of increasing roughness called `mirror', `mist' and `hackle' which correspond to progressive crack acceleration from subcritical to dynamic propagation (Johnson and Holloway 1966).
The spread of AFM investigations in the '90s has shown that surface roughness in the `mirror' region is of the order of a fraction of nm (RMS) when measured on micrometer size images, and that it is almost comparable with the roughness of melt surfaces and with the best levels of surface finishing by polish treatments (cf.\ Rädlein and Frischat 1997; Arribart and Abriou 2000).

The dependence of the RMS roughness of fracture surfaces in the mirror region on the nature of glass was extensively studied to understand the relation between the heterogeneities in the glass composition and the crack path through the glass (Gupta \etal 2000; Wiederhorn \etal 2007). Other investigators have explored the scaling properties of the fracture surfaces (Bouchaud 1997) and the dependence of the self-affine exponents and the characteristic cutoff length on the velocity of crack propagation (Bonamy \etal 2006).

The extensive development of dynamical models for the crack front line in a heterogeneous material (cf.\ Ramanathan \etal 1997), have made the investigation of the scaling behaviour of glass fracture surfaces a very interesting tool for probing the physics of crack propagation. A particularly interesting aspect is the correlation between the position of some cutoff lengths in the scaling properties with the size of the process zone as observed in several different materials (Bonamy \etal 2006). These observations support the idea that a generalised process zone behaviour can affect scales of the order of a few tens of nanometres during slow crack propagation in silica glass, the size of the process zone being an increasing function of the decreasing crack propagation velocity.

% S. Ramanathan, D. Ertas, and D. S. Fisher, Phys. Rev. Lett. 79, 873 (1997).
% S. Ramanathan and D. S. Fisher, Phys. Rev. Lett. 79, 877 (1997).

A useful complementary tool to study the damage mechanisms induced by crack propagation is the analysis of the mismatches in the recombination of the two opposite fracture surfaces. Guin \etal 2004 showed that it is possible to match down to nanometre scale opposite fracture faces of a soda-lime glass broken in water. The fine comparison on a few sections normal to the crack surface showed no evidence of damage cavities in the bulk of the glass sample (cf.\ also Fett \etal 2008). However, the determination of the remnants of damage processes on {\it post-mortem} (unstressed) fracture surfaces in glass requires extreme care since the relevant information is at the nanometre scale where the metrologic capacities of AFM imaging are affected by several delicate problems such as drifts, feedback effects, noise and tip shape artifacts (Flemming \etal 2005). The accurate treatment of these kinds of problems will benefit from the present development of digital image correlation techniques (Hild and Roux 2006) and requires the application of advanced statistical estimation tools.

As discussed in section \ref{sec:threshold}, {\it post-mortem} AFM investigations and recombination techniques have also led to important insights into the nature of the propagation threshold and the mechanisms of crack arrest and aging. For example, Guin \etal 2005 could measure the extent of the dissolution of the crack surfaces when a crack in soda-lime glass is arrested and aged near the propagation threshold in a basic solution. The work of Watanabe \etal 1994 on the evolution of the fracture surfaces of a soda-lime glass after fracture in vacuum has provided important insight into the time scales of the corrosion mechanisms caused by the interaction with the moist atmosphere.

The UHV AFM investigations of the fracture surfaces of different kinds of glasses with molecular resolution (Frischat \etal 2004) have provided an intimate view of the complex nanostructures
in multi-component glasses and their role in the slow crack propagation processes. These kinds of measurements are very promising for relating the  structural measurements in bulk glasses (cf.\ Wright 2008) with the molecular simulations of the peculiar structure of the glass fracture surfaces (Du and Cormack 2005; Roder \etal 2001).

Many techniques for the high resolution characterisation of the structural and chemical properties of glass surfaces are being developed, which are out of the scope of the present review. I will just cite here some relevant investigation of the hydroxylation behaviour of fracture surfaces (Souza and Pantano 2002) since the study of the water adsorption sites and of the time scales of the hydroxylation mechanisms are of particular interest for understanding the high energy of the glass fracture surfaces (Leed and Pantano 2003) and modelling their wetting properties, which were shown to have a paramount role in the corrosion mechanisms in moist environment (cf.\ section \ref{sec:IonMigration}).

\section{Perspectives}
\label{sec:Perspectives}

After almost one hundred years from the initial insights into the nature of brittle crack propagation in glasses we finally feel the flavour of a turning point in the comprehension of the basic mechanisms of slow crack propagation in glasses. The kernel of the difficulty turned out to be that there are many actors playing in the opera, most of them acting at the nanoscale in a deeply interrelated manner. The amorphous nature of glass has deprived the scientists of most of their typical modelling and characterisation tools, and the exceptional homogeneity of glass relegated all the relevant phenomena to the nanoscale, where the chemical nature of surfaces becomes most influent and the strong confinement makes the macroscopic thermodynamics fail. On the other hand, the disordered structure of both the glass network and the reactive molecules (such as water) tend to appear ordered at the nanoscale, a fascinating order that is often specific to each of the many glass compositions.

The turning point stems from the progressive adaptation to the nanoscale of both the investigation tools and the simulation techniques, along with the increasing comprehension of the physics of the nanoworld. Different kinds of local scanning probes have allowed the visualisation of the space and time scales of the physico-chemical mechanisms during or after crack propagation. Several structural and compositional analyses are also increasing their resolution and reducing the severity of their condition of use. The most promising developments concern the combination of different complementary probes on a single setup (such as AFM+microRaman techniques). At the same time, the gap between simulation and experience is also being drastically reduced. On one hand, both molecular dynamics and first principle simulations permit the modelling of increasing volumes of matter and increasing time durations (which remain however the limiting factor). On the other hand, several experiments under UHV conditions are exploring physical conditions that approach those being simulated (by exploring for example the interaction of few water molecules with the glass fracture surface).

However, when trying to model these new exciting investigations, we should not forget all the ingredients of a sound mechanical modelling and the subtle details of glass surface chemistry.
The crack tip and process zone phenomena should be adequately separated from the structural behaviour of the glass sample in order to tackle correctly
the issues related to the management of residual stresses, to surface finishing, and to the effects of the conditioning treatments on the population of  surface flaws and nucleation sites.
A major difficulty highlighted in this review is the subtle difference between the properties of crack propagation or surface corrosion processes in liquid solutions and under moist atmosphere. While the first case corresponds to an overall controlled environment (at least for slow crack propagation), in the second case the local environment at the crack tip or at the glass fracture and external surfaces is determined by the evolution in time of a very confined liquid condensation. Carbon dioxide in the atmosphere was also shown to be a major actor in the evolution of the different chemical reactions. It is thus of fundamental importance to dispose of a high degree of control of the atmosphere during the investigations, and to repeat the experiences under pure mixtures of nitrogen and humidity.

This review is not exhaustive, but hopefully sufficient analysis has been provided in order to promote a deep reflection on the relevance of the different actors. This reflection should guide to the identification of the space and time scales relevant for the action of the different physical and chemical processes in the close neighbourhood of the crack tip and crack surfaces as a function of the different glass compositions and controlled environments. This analysis should always be complemented by the estimation of the energetic contribution of each individual mechanism. Special attention should be paid to the systematic complete characterisation of the glass samples used, including their detailed composition, structural and mechanical properties and to the definition of the thermal history as well as the details of the surface finishing techniques. The reproducibility of the results in different laboratories will be key point for a real understanding.

We should not forget that understanding the nature of the crack propagation mechanisms will help in solving just part of the problem of glass strength. This line of research should be combined with a comprehension of the flaw nucleation mechanisms starting from different kinds of defects within the glass and with the development of techniques making glass objects more tolerant to the presence of such flaws. Strong efforts in these directions are being invested by both the academic and industrial communities in order to make glass a promising material for advanced applications in the XXI century.

\ack

This work was motivated by the reflections on the animation of the Technical Committee TC09 on Glass Nanomechanics in the International Commission on Glass and by the organisation of a workshop financed by the EFONGA project to be held in Montpellier on February 22-25, 2009. The research activity was financed by the ANR project CORCOSIL (BLAN07-3\_196000). I thank all the TC09 and TC19 members for nice discussions as well as the colleagues of the Glass group in Montpellier and the partners of our ANR project in CEA-Saclay and Université Lyon I. A special thanks to S. Wiederhorn and M. George for critical reading of the manuscript, to I. Campbell for carefully proof reading it, and to C. Marlière and F. Célarié for initially developing the Nanomechanics activity in Montpellier.

\section*{References}

%\begin{thebibliography}{1}

\begin{harvard}

%\bibitem{Arribart2000}
\item[]
Arribart H and Abriou D 2000
%Ten years of atomic force microscopy in glass research
{\it Ceramics-Silikáty} {\bf 44} 121--8 % OK

%\bibitem{BandoTomoz1984}
\item[]
Bando Y, Ito S and Tomozawa M 1984
%Direct Observation of Crack Tip Geometry of $SiO_2$ Glass by High-Resolution Electron Microscopy
{\it J.\ Am.\ Ceram.\ Soc} {\bf 67} C36--7

%\bibitem{Barenblatt1962}
\item[]
Barenblatt G I 1962
%The mathematical theory of equilibrium cracks in brittle fracture.
in {\it Advances in Applied Mechanics} vol 7 (New York: Academic Press) pp 55--129

%\bibitem{Barquins1997}
\item[]
Barquins M and Ciccotti M 1997
%On the kinetics of peeling of an adhesive tape under a constant imposed load
{\it Int.\ J.\ Adhes.\ Adhes.} {\bf 17} 65--8 % OK

%\bibitem{Bazant1979}
\item[]
Bazant Z P and Estenssoro L F 1979
{\it Int.\ J.\ Solids Struct.} {\bf 15} 405--26

%\bibitem{Bouchaud1997}
\item[]
Bouchaud E 1997
%Scaling properties of cracks
{\it J.\ Phys.: Condens.\ Matter} {\bf 9} 4319--44

%\bibitem{Bonamy2006}
\item[]
Bonamy D, Ponson L, Prades S, Bouchaud E and Guillot C 2006
%Scaling Exponents for Fracture Surfaces in Homogeneous Glass and Glassy Ceramics
{\it Phys.\ Rev.\ Lett.} {\bf 97} 135504

%\bibitem{Broberg2003}
\item[]
Broberg K B 2003
%The many scales in fracture mechanics
{\it Strength, Fracture and Complexity} {\bf 1} 31--8

%\bibitem{Bunker1994}
\item[]
Bunker B C 1994
%Molecular mechanisms for corrosion in silica and silicate glasses
{\it J.\ Non-Cryst.\ Solids} {\bf 179} 300--8 %(OK)

%\bibitem{Celarie2007}
\item[]
Célarié F, Ciccotti M and Marlière C 2007
%Stress-enhanced ion diffusion at the vicinity of a crack tip as evidenced by atomic force microscopy in silicate glasses
{\it J.\ Non-Cryst.\ Solids} {\bf 353} 51--68

%\bibitem{Celarie2003}
\item[]
Célarié F, Prades S, Bonamy D, Ferrero L, Bouchaud E, Guillot C and Marlière C 2003
{\it Phys.\ Rev.\ Lett.} {\bf 90} 075504

%\bibitem{CH1961}
\item[]
Charles R J and Hillig W B 1962 in {\it Symp.\ on Mechanical Strength of Glass and Ways of Improving It} Florence, Italy, September 25-29, 1961 (Charleroi, Belgium: Union Scientiftque Continentale du Verre) pp 511--27

%\bibitem{Ciccotti2008}
\item[]
Ciccotti M, George M, Ranieri V, Wondraczek L and Marlière C 2008
%Dynamic condensation of water at crack tips in fused silica glass.
{\it J.\ Non-Cryst.\ Solids.} {\bf 354} 564--8

%\bibitem{Ciccotti2004}
\item[]
Ciccotti M, Giorgini B, Vallet D and Barquins M 2004
%Complex dynamics in the peeling of an adhesive tape
{\it Int.\ J.\ Adhes.\ Adhes.} {\bf 24} 143--51 % OK

%\bibitem{Conr2008}
\item[]
Conradt R 2008
%Chemical Durability of Oxide Glasses in Aqueous Solutions: A Review
{\it J.\ Am.\ Ceram.\ Soc.} {\bf 91} 728--35 % OK

%\bibitem{Crichton1999}
\item[]
Crichton S N, Tomozawa M, Hayden J S, Suratwala T I, Campbell J H 1999
%Subcritical Crack Growth in a Phosphate Laser Glass
{\it J.\ Am.\ Ceram.\ Soc.} {\bf 82} 3097--104

%\bibitem{Davidge1973}
\item[]
Davidge R W, McLaren J R and Tappin G 1973
%Strength Probability Time (SPT) Relationships in Ceramics
{\it J.\ Mater.\ Sci} {\bf 8} 1699--709

%\bibitem{DelGa2007}
\item[]
Del Gaudio P, Behrens H and Deubener J 2007
%Viscosity and glass transition temperature of hydrous float glass
{\it J.\ Non-Cryst.\ Solids} {\bf 353} 223--36

%\bibitem{Dim}
\item[]
Dimitrov A, Buchholz F G and Schnack E 2006
%On Three-dimensional Effects in Propagation of Surface-breaking Cracks
{\it Comp.\ Model.\ Eng.\ Sci.} {\bf 12} 1--25

%\bibitem{Doremus1975}
\item[]
Doremus R H 1975
%Interdiffusion of hydrogen and alkali ions in a glass surface
{\it J.\ Non-Cryst.\ Solids} {\bf 19} 137--44

%\bibitem{Dor2002}
\item[]
Doremus R H 2002
{\it Diffusion of Reactive Molecules in Solids and Melts} (New York: John Wiley \& Sons)

%\bibitem{Cormack2005}
\item[]
Du J and Cormack A N 2005
%Molecular Dynamics Simulation of the Structure and Hydroxylation of Silica Glass Surfaces
{\it J.\ Am.\ Ceram.\ Soc.} {\bf 88} 2532-–39

%\bibitem{Dug1960}
\item[]
Dugdale D S 1960
%Yielding of steel sheets containing slits
{\it J.\ Mech.\ Phys.\ Solids} {\bf 8} 100--4

%\bibitem{Etter2008}
\item[]
Etter S, Despetis F and Etienne P 2008
%Sub-critical crack growth in some phosphate glasses
{\it J.\ Non-Cryst.\ Solids} {\bf 354} 580--6

%\bibitem{Evans1972}
\item[]
Evans A G 1972
%A Method for Evaluating the Time-Dependent Failure Characteristics of Brittle Materials—And Its Applications to Polycrystalline Alumina
{\it J.\ Mater.\ Sci.} {\bf 7} 1137-–46

%\bibitem{Evans1974}
\item[]
Evans A G and Wiederhorn S M 1974
%Proof testing of ceramic materials - an analytical basis for failure prediction
{\it Int.\ J.\ Fract.} {\bf 10} 379--92

%\bibitem{Fett2005a}
\item[]
Fett T, Guin J P and Wiederhorn S M 2005a
%Interpretation of effects at the static fatigue limit of soda-lime-silicate glass
{\it Eng.\ Fract.\ Mech.} {\bf 72} 2774–-91

%\bibitem{Fett2005b}
\item[]
Fett T, Guin J P and Wiederhorn S M 2005b
%Stresses in ion-exchange layers of soda-lime-silicate glass
{\it Fatigue Fract.\ Eng.\ Mater.\ Struct.} {\bf 28} 507–-14 % OK

%\bibitem{Fett2008}
\item[]
Fett T, Rizzi G, Creek D, Wagner S, Guin J P, López-Cepero J M, and Wiederhorn S M 2008
{\it Phys.\ Rev.\ B} {\bf 77} 174110

%\bibitem{Fineberg1999}
\item[]
Fineberg J and Marder M 1999
%Instability in dynamic fracture
{\it Phys.\ Rep.} {\bf 313} 1--108 % OK

%\bibitem{Flemming2005}
\item[]
Flemming M, Roder K and Duparré A 2005
%Scanning force microscopy for optical surface metrology.
{\it Proc. SPIE}
% - The International Society for Optical Engineering, USA
{\bf 5965} 59650A:1--10

%\bibitem{Freim1975}
\item[]
Freiman S W 1975
%Effect of Straight-Chain Alkanes on Crack Propagation in Glass
{\it J.\ Am.\ Ceram.\ Soc.} {\bf 58} 339--41

%\bibitem{Freund1990}
\item[]
Freund L B 1990 {\it Dynamic Fracture Mechanics} (Cambridge: Cambridge University Press)

%\bibitem{Frischat2004}
\item[]
Frischat G H, Poggemann J F and Heide G 2004
%Nanostructure and atomic structure of glass seen by atomic force microscopy
{\it J.\ Non-Cryst.\ Solids} {\bf 345} 197-–202

%\bibitem{Gehrke1991}
\item[]
Gehrke E, Ullner C and Mahnert M 1991
%Fatigue limit and crack arrest in alkalicontaining silicate glasses
{\it J.\ Mater.\ Sci.} {\bf 26} 5445--55 % OK \ddot{a}

%\bibitem{Glasstone1941}
\item[]
Glasstone S, Leidler K J and Eyring H 1941
{\it The Theory of Rate Processes} (New York: McGraw-Hill)

%\bibitem{Gorsky1935}
\item[]
Gorsky W S 1935
%Theorie des elastichen Nachwirkung in ungeordneten Mischkristallen
%Physikalische Zeitschrift der Sowjetunion
{\it Phys.\ Zeit.\ Sowjet.} {\bf 8} 457--71

%\bibitem{Grif1921}
\item[]
Griffith A A 1921
%The phenomena of rupture and flow in solids
{\it Phil.\ Trans.\ Roy.\ Soc.\ London} {\bf A221} 163--198

%\bibitem{GRIMALDI2008}
\item[]
Grimaldi A, George M, Pallares G, Marlière C and Ciccotti M 2008
%The crack tip: a nanolab for studying confined liquids
{\it Phys.\ Rev.\ Lett.} {\bf 100} 165505

%\bibitem{Gui1996}
\item[]
Guilloteau E, Charrue H and Creuzet F 1996
{\it Europhys.\ Lett.} {\bf 34} 549--53

%\bibitem{Guin2004}
\item[]
Guin J P and Wiederhorn S M 2004
%Fracture of Silicate Glasses: Ductile or Brittle?
{\it Phys.\ Rev.\ Lett.} {\bf 92} 215502

%\bibitem{Guin2005}
\item[]
Guin J P and Wiederhorn S M 2005
%Crack-Tip Structure in Soda–Lime–Silicate Glass
{\it J.\ Am.\ Ceram.\ Soc.} {\bf 88} 652--59

%\bibitem{Gupta2000}
\item[]
Gupta P K, Inniss D, Kurkjian C R and Zhong Q 2000
%Nanoscale roughness of oxide glass surfaces
{\it J.\ Non-Cryst.\ Solids} {\bf 262} 200--6

%\bibitem{Gy2003}
\item[]
Gy R 2003
%Stress corrosion of silicate glass: a review
{\it J.\ Non-Cryst.\ Solids} {\bf 316} 1--11

%\bibitem{Gy2008}
\item[]
Gy R 2008
%Ion exchange for glass strengthening
{\it Mater.\ Sci.\ Eng.\ B} {\bf 149} 159--65

%\bibitem{HanTomoz1989}
\item[]
Han W T and Tomozawa M 1989
%Mechanism of Mechanical Strength Increase of Soda-lime Glass by Aging
{\it J.\ Am.\ Ceram.\ Soc} {\bf 72} 1837--43

%\bibitem{HanTomoz1989}
%\item[]
%Henaux S and Creuzet F 2000
%Crack Tip Morphology of Slowly Growing Cracks in Glass
%{\it J.\ Am.\ Ceram.\ Soc} {\bf 83} 415--17

%\bibitem{Roux2006}
\item[]
Hild F and Roux S 2006
%Digital image correlation: from displacement measurement to identification of elastic properties - a review.
{\it Strain} {\bf 42} 69--80

%\bibitem{Iler1979}
\item[]
Iler R K 1979
{\it The Chemistry of Silica} (New York: John Wiley and Sons)

%\bibitem{Irwin1960}
\item[]
Irwin G R 1960
%Plastic zone near a crack and fracture toughness.
in {\it Mechanical and Metallurgical Behavior of Sheet Material (Proc.\ 7th Sagamore Ordnance Materials Research Conf., Syracuse)} (Arlington, Virginia: ASTIA) pp IV.63--71
% Voir Sanford
% Proceedings of the Seventh Sagamore Ordnance Materials Research Conference, 1960, pp. IV-63–IV-78.

%\bibitem{Irwin1957}
\item[]
Irwin G R and Washington D C 1957
%Analysis of stresses and strains near the end of a crack travesing a plate
{\it J.\ Appl.\ Mech.} {\bf 54} 361--4

%\bibitem{ItoTomoz1982}
\item[]
Ito S and Tomozawa M 1982
%Crack Blunting of High-Silica Glass
{\it J.\ Am.\ Ceram.\ Soc.} {\bf 65} 368--71

%\bibitem{Janssen1974}
\item[]
Janssen C 1974
%Specimen for fracture mechanics studies in glass
in {\it Proc. 10th Int. Cong. on Glass (Kyoto)}
%July 8-13, 1974, Kyoto, Japan
(Tokyo, Japan: Ceramic Society of Japan) pp 10.23--10.30

%\bibitem{Johnson1966}
\item[]
Johnson J W and Holloway D G 1966
{\it Phil.\ Mag.} {\bf 14} 731--43 % issue 130

%\bibitem{Hand2000}
Kingston JGR and Hand RJ 2000
%Compositional effects on the fracture behaviour of alkali-silicate glasses
{\it Fatigue Fract.\ Engng.\ Mat.\ Struct.} {\bf 23} 685--90

%\bibitem{Kurkjian2000}
Kurkjian CR, Gupta PK, Brow RK and Lower N 2003
%The intrinsic strength and fatigue of oxide glasses
{\it J.\ Non-Cryst.\ Sol.} {\bf 316} 114--24

%\bibitem{Tomoz2006}
\item[]
Koike A and Tomozawa M 2006
%Fictive temperature dependence of subcritical crack growth rate of normal glass and anomalous glass
{\it J.\ Non-Cryst.\ Solids} {\bf 352} 5522--30

%\bibitem{Lanf1979}
\item[]
Lanford W A, Davis K, Lamarche P, Laursen T, Groleau R, Doremus R H 1979
%Hydration of soda-lime glass
{\it J.\ Non-Cryst.\ Sol.} {\bf 33} 249–66

%\bibitem{Langford1991}
\item[]
Langford S C, Jensen L C, Dickinson J T and Pederson L R 1991
%Alkali emission accompanying fracture of sodium silicate glasses
{\it J.\ Mat.\ Res.} {\bf 6} 1358--68

%\bibitem{Lawn1983}
\item[]
Lawn B R 1983
%Physics of Fracture
{\it J.\ Am.\ Ceram.\ Soc.} {\bf 66} 83--91

%\bibitem{Lawn1993}
\item[]
Lawn B R 1993 {\it Fracture of Brittle Solids} 2nd ed (Cambridge: Cambridge University Press)

%\bibitem{Lawn1980}
\item[]
Lawn B R, Hockey B J and Wiederhorn S M 1980
%Atomically sharp cracks in brittle solids: an electron microscopy study.
{\it J.\ Mat.\ Sci.} {\bf 15} 1207--23

%\bibitem{Lawn1985}
\item[]
Lawn B R, Jakus K and Gonzales A C 1985
%Sharp vs Blunt Crack Hypotheses in the Strength of Glass: A Critical Study Using Indentation Flaws
{\it J.\ Am.\ Ceram.\ Soc.} {\bf 68} 25--34

%\bibitem{Pantano2003}
\item[]
Leed E A and Pantano C G 2003
%Computer modeling of water adsorption on silica and silicate glass fracture surfaces
{\it J.\ Non-Cryst.\ Solids} {\bf 325} 48--60

%\bibitem{Tomoz1995}
\item[]
Li H, Agarwal A and Tomozawa M 1995
%Effect of fictive temperature on dynamic fatigue behavior of silica and soda-lime glass
{\it J.\ Am.\ Ceram.\ Soc.} {\bf 78} 1393--6

%\bibitem{Lopez2007}
\item[]
López-Cepero J M, Wiederhorn S M, Fett T and Guin J P 2007
%Do plastic zones form at crack tips in silicate glasses?
{\it Int.\ J.\ Mat.\ Res.} {\bf 98} 1170--6

%\bibitem{Marsh1964a}
\item[]
Marsh D M 1964a
{\it Proc.\ R.\ Soc.\ London Ser.\ A} {\bf 279} 420--74

%\bibitem{Marsh1964b}
\item[]
Marsh D M 1964b
{\it Proc.\ R.\ Soc.\ London Ser.\ A} {\bf 282} 33--43

%\bibitem{Maugis1985}
\item[]
Maugis D 1985
%Review Subcritical crack growth, surface energy, fracture toughness, stick-slip and embrittlement
{\it J.\ Mater.\ Sci.} {\bf 20} 3041--73

%\bibitem{Maug1992}
\item[]
Maugis D 1992
{\it Eng.\ Fract.\ Mech.} {\bf 43} 217--55

%\bibitem{Michalske1977}
\item[]
Michalske T A 1977
%The Stress Corrosion Limit: Its Measurement and Implications
in {\it Fracture Mechanics of Ceramics} vol 5
%Surface Flaws, Statistics, and Microcracking.
ed R C Bradt, A G Evans \etal
%, D. P. H. Hasselman, and F. F. Lange.
(New York: Plenum Press) pp 277--89

%\bibitem{Mich1984}
\item[]
Michalske T A and Bunker B C 1984
%Slow fracture model based on strained silicate structures
{\it J.\ Appl.\ Phys} {\bf 56} 2686--93 % OK

%\bibitem{Mich1987}
\item[]
Michalske T A and Bunker B C 1987
%Steric Effects in Stress Corrosion Fracture of Glass
{\it J.\ Am.\ Ceram.\ Soc.} {\bf 70} 780--4 % OK

%\bibitem{Mich1989}
\item[]
Michalske T A and Bunker B C 1989
%Effect of Surface Stress on Stress Corrosion of Silicate Glass
in {\it Advances in Fracture Research} vol 6
ed K Salama, K Ravi-Chandler \etal
%, D M R Tablin and P Ramao Rao, Eds.
(New York: Pergamon Press) pp 3689--99

%\bibitem{Mich1993}
\item[]
Michalske T A and Bunker B C 1993
%A Chemical Kinetics Model for Glass Fracture
{\it J.\ Am.\ Ceram.\ Soc.} {\bf 76} 2613--8 % OK Cyclosiloxanes

%\bibitem{Michaslke1980}
\item[]
Michalske T A and Frechette V D 1980
%Dynamic Effects of Liquids on Crack Growth Leading to Catastrophic Failure in Glass
{\it J.\ Am.\ Ceram.\ Soc.} {\bf 63} 603--9

%\bibitem{Michaslke1983}
\item[]
Michalske T A and Freiman S W 1983
%A Molecular Mechanism for Stress Corrosion in Vitreous Silica
{\it J.\ Am.\ Ceram.\ Soc.} {\bf 66} 284--8

%\bibitem{Mich1985}
\item[]
Michalske T A and Fuller E R 1985
%Closure and repropagation of heald cracks in silicate glasses
{\it J.\ Am.\ Ceram.\ Soc.} {\bf 68} 586--90

%\bibitem{Mould1967}
\item[]
Mould RE 1967
%"The Strength of Inorganic Glasses,"
in {\it Fundamental Phenomena in the Materials Sciences} vol 4
%Fracture of Metals Polymers and Glasses,
ed L J Bonis, J J Duga and J.J. Gilman
(New York: Plenum Press) pp 119--49

%\bibitem{Mura1996}
\item[]
Muraoka M and Abé H 1996
%Subcritical Crack Growth in Silica Optical Fibers in a Wide Range of Crack Velocities
{\it J.\ Am.\ Ceram.\ Soc.} {\bf 79} 51--7

%\bibitem{Nghiem1998}
\item[]
Nghiem B 1998
PhD Thesis. Université Paris-VI, France

%\bibitem{Nogami1984}
\item[]
Nogami M and Tomozawa M 1984
%Effect of Stress on Water Diffusion in Silica Glass
{\it J.\ Am.\ Ceram.\ Soc.} {\bf 67} 151--4

%\bibitem{Orowan1955}
\item[]
Orowan E 1955
%Energy criteria of fracture
{\it Weld.\ J.\ Res.\ Suppl.} {\bf 34} S157--160 % OK ex Sannford

%\bibitem{Pant1985}
\item[]
Pantano C G 1985
in {\it Strength of Inorganic Glass (Nato Conference Series VI)} ed C R Kurkjian (New York: Plenum Press) pp 37--66

%\bibitem{Tomoz1997}
\item[]
Peng Y L, Tomozawa M and Blanchet T A 1997
%Tensile stress-acceleration of the surface structural relaxation of SiO$_2$ optical fibers
{\it J.\ Non-Cryst.\ Solids} {\bf 222} 376--82

%\bibitem{Pez2008}
\item[]
Pezzotti G, Leto A and Porporati AA 2008
%Visualization of microscopic stress fields in silica glass in the scanning electron microscope
{\it J.\ Ceram.\ Soc.\ Japan} {\bf 116} 869--74

%\bibitem{Pollet1977}
\item[]
Pollet J C and Burns S J 1977
%An analysis of slow crack propagation data in PMMA and brittle materials
{\it Int.\ J.\ Fract.} {\bf 13} 775--86 % OK

%\bibitem{Prades2005}
\item[]
Prades S, Bonamy D, Dalmas D, Bouchaud E, Guillot C 2005
{\it Int.\ J.\ Solids Struct.} {\bf 42} 637--45

%\bibitem{Preston1942}
\item[]
Preston FW 1942
%The Mechanical Properties of Glass
{\it J.\ Appl.\ Phys.} {\bf 13} 623--34

%\bibitem{Proctor1962}
\item[]
Proctor R 1962
%The Effects of Hydrofluoric Acid Etching on the Strength of Glasses
{\it Phys.\ Chem.\ Glasses} {\bf 3} 7--27 %OK but miss

%\bibitem{Frischat1997}
\item[]
Rädlein E and Frischat G H 1997
%Atomic force microscopy as a tool to correlate nanostructure to properties of glasses
{\it J.\ Non-Cryst.\ Solids} {\bf 222} 69--82

%\bibitem{Ramanathan1997}
\item[]
Ramanathan S, Ertas D and Fisher D S 1997
{\it Phys.\ Rev.\ Lett.} {\bf 79} 873--6
% S. Ramanathan and D. S. Fisher, Phys. Rev. Lett. 79, 877 (1997).

%\bibitem{Rice1968}
\item[]
Rice J R 1968
%A path independent integral and approximate analysis of strain concentration by nothces and cracks
{\it J.\ Appl.\ Mech.} {\bf 35} 379--86

%\bibitem{Rice1978}
\item[]
Rice J R 1978
%Thermodynamics of the quasi-static growth of Griffith cracks
{\it J.\ Mech.\ Phys.\ Solids} {\bf 26} 61--78

%\bibitem{Rizkalla1992}
\item[]
Rizkalla A S, Jones D W and Sutow E J 1992
%Effect of Nonbridging Oxygens on the Fracture Toughness of Synthesized Glasses
{\it Br.\ Ceram.\ Trans.\ J.} {\bf 91} 12--15

%\bibitem{Kob2001}
\item[]
Roder A, Kob W and Binder K 2001
%Structure and Dynamics of amorphous Silica Surfaces
{\it J.\ Chem.\ Phys.} {\bf 114} 7602--14

%\bibitem{Pantano2002}
\item[]
Souza A S and Pantano C G 2002
%Hydroxylation and Dehydroxylation Behavior of Silica Glass Fracture Surfaces
{\it J.\ Am.\ Ceram.\ Soc.} {\bf 85} 1499--504

%\bibitem{Srawley1967}
\item[]
Srawley J E and Gross B 1967
%Stress Intensity Factors for Crackline-Loaded Edge-Crack Specimens
{\it Mater.\ Res.\ Std.} {\bf 7} 155--62

%\bibitem{Stav1983}
\item[]
Stavrinidis B and Holloway D G 1983
%Crack healing in glasses
{\it Phys.\ Chem.\ Glasses} {\bf 24} 19--25

%\bibitem{Stuke2006}
\item[]
Stuke A, Behrens H, Schmidt B C and Dupré R 2006
%H$_2$O speciation in float glass and soda lime silica glass
{\it Chem.\ Geol.} {\bf 229} 64--77

%\bibitem{Surat2000}
%\item[]
%Suratwala T I, Steele R A, Wilke G D, Campbell J H and Takeuchi K 2000
%Effects of OH content, water vapor pressure, and temperature on sub-critical crack growth in phosphate glass.
%{\it J.\ Non-Cryst.\ Solids} {\bf 263} 213--27

%\bibitem{Taylor1949}
\item[]
Taylor E W 1949
%Plastic Deformation of optical glass
{\it Nature} {\bf 163} 323--3

%\bibitem{Thom1971}
\item[]
Thomson R, Hsieh C and Rana V 1971
%Lattice trapping of fracture cracks
{\it J.\ Appl.\ Phys.} {\bf 42} 3154--60

%\bibitem{Tool1946}
\item[]
Tool A Q 1946
%Relation between inelastic deformability and thermal expansion of glass in its annealing range
{\it J.\ Am.\ Ceram.\ Soc.} {\bf 29} 240--53

%\bibitem{Tomoz1996}
\item[]
Tomozawa M 1996
%Fracture of Glasses
{\it Annu.\ Rev.\ Mater.\ Sci.} {\bf 26} 43--74

%\bibitem{Tomoz1991}
\item[]
Tomozawa M, Han W T and Lanford W A 1991
%Fracture of Glasses
{\it J.\ Am.\ Ceram.\ Soc.} {\bf 74} 2573--6

%\bibitem{Varner2006}
%\item[]
%Varner J 2006
%{\it Strength and fracture mechanics of glass}
%in: {\it ICG Advanced Course 2006: Strength of Glass, Basic and Test Procedures} (Offenbach, Germany: Verlag DGG) pp 23--37

%\bibitem{Cook2008}
\item[]
Vaudin MD, Gerbig YB, Stranick SJ and Cook RF 2008
%Comparison of nanoscale measurements of strain and stress using electron back scattered diffraction and confocal Raman microscopy
{\it Appl.\ Phys.\ Lett.} {\bf 93} 193116
% Cook R F 2008 Private communication

%\bibitem{Vernaz1978}
\item[]
Vernaz E 1978
%Influence de la composition sur la ténacité des verres,
PhD thesis, Université Montpellier 2, France

%\bibitem{Vernaz1978}
\item[]
Vernaz E, Larché F and Zarzycki J 1980
%Fracture toughness-composition relationship in some binary and ternary glass systems
{\it J.\ Non-Cryst.\ Solids} {\bf 37} 359--65

%\bibitem{WakabTomoz1989}
%\item[]
%Wakabayashi H and Tomozawa M 1989
%Diffusion of Water into Silica Glass at low Temperature
%{\it J.\ Am.\ Ceram.\ Soc.} {\bf 72} 1850--5

%\bibitem{Wan1990a}
\item[]
Wan K T, Aimard N. Lathabai S, Horn R G and Lawn B R 1990
{\it J.\ Mat.\ Res.} {\bf 5} 172--82 %(Lawn p121)

%\bibitem{Wan1990b}
\item[]
Wan K T, Lathabai S and Lawn B R 1990
% Crack velocity functions and threshold in brittle solids
{\it J.\ European Ceram.\ Soc.} {\bf 6} 259--68 %(Lawn p120)

%\bibitem{Wata1994}
\item[]
Watanabe Y, Nakamura Y, Dickinson J T and Langford S C 1994
%Changes in air exposed fracture surfaces of silicate glasses observed by atomic force microscopy
{\it J.\ Non-Cryst.\ Solids} {\bf 177} 9--25

%\bibitem{Weber1964}
\item[]
Weber N and Goldstein M 1964
%Stress-induced migration and partial molar volume of sodium ions in glass
{\it J Chem Phys} {\bf 41} 2898--901

%\bibitem{Wied1967}
\item[]
Wiederhorn S M 1967
%Influence of Water Vapor on Crack Propagation in Soda-Lime Glass.
{\it J.\ Am.\ Ceram.\ Soc.} {\bf 50} 407--14

%\bibitem{Wied1967}
\item[]
Wiederhorn S M 1969
%Fracture Surface Energy of Glass
{\it J.\ Am.\ Ceram.\ Soc.} {\bf 52} 99--105

%\bibitem{Wieder1970}
\item[]
Wiederhorn S M and Bolz L H 1970
%Stress Corrosion and Static Fatigue of Glass
{\it J.\ Am.\ Ceram.\ Soc.} {\bf 53} 543--8 %OK

%\bibitem{Wiederhorn1982}
\item[]
Wiederhorn S M, Freiman S W, Fuller E R Jr and Simmons C J 1982
%Effect of Water and Other Dielectrics on Crack Growth
{\it J.\ Mater.\ Sci.} {\bf 17} 3460--78

%\bibitem{Wie1973}
\item[]
Wiederhorn S M and Johnson H 1973
%Effect of Electrolyte pH on Crack Propagation in Glass
{\it J.\ Am.\ Ceram.\ Soc.} {\bf 56}  192--7 % OK

%\bibitem{Wie1974}
\item[]
Wiederhorn S M, Johnson H, Diness A M and Heuer A H 1974
%Fracture of Glass in Vacuum
{\it J.\ Am.\ Ceram.\ Soc.} {\bf 57} 336--41 % OK

%\bibitem{Wiederhorn2007}
\item[]
Wiederhorn S M, López-Cepero J M, Wallace J, Guin J P and Fett T 2007
%Roughness of glass surfaces formed by sub-critical crack growth
{\it J.\ Non-Cryst.\ Solids} {\bf 353} 1582--91

%\bibitem{Wondra2006}
\item[]
Wondraczek L, Ciccotti M, Dittmar A, Oelgardt C, Célarié F and Marlière C 2006
%Real-time observation of non-equilibrium liquid condensate confined at tensile crack tips in oxide glasses
{\it J.\ Am.\ Ceram.\ Soc.} {\bf 89} 746--9

%\bibitem{Wright2008}
\item[]
Wright A C
% Longer range order in single component network glasses?
{\it Phys. Chem. Glasses} {\bf 49} 103--117

%\bibitem{Zarzycki2005}
\item[]
Zarzycki J 1991 {\it Glasses and the Vitreous State} (Cambridge: Cambridge University Press)

%\bibitem{Pez2007}
\item[]
Zhu W, Porporati A A, Matsutani A, Lama N and Pezzotti G 2007
%Spatially resolved crack-tip stress analysis in semiconductor by cathodoluminescence piezospectroscopy
{\it J.\ Appl.\ Phys.} {\bf 101} 103531 % 1-12

%\end{thebibliography}
\end{harvard}

\end{document}